\documentclass[]{emulateapj}

\def\figdir{./}
\newcommand{\ltsima}{\mbox{$\; \buildrel < \over \sim \;$}}
\def\simlt{\lower.5ex\hbox{\ltsima}}
\def\gtsima{$\; \buildrel > \over \sim \;$}
\def\simgt{\lower.5ex\hbox{\gtsima}}

\def\eq{equation}

\def\HI{\mbox{\ion{H}{1}}}
\def\HII{\mbox{\ion{H}{2}}}

\def\GII{\mbox{\ion{He}{2}}}

\def\figname#1{\figdir/#1}

\def\eq{equation}
\def\fig{Fig.}

\def\ie{{\it i.e.}}
\def\eg{{\it e.g.}}
\def\ltsima{$\; \buildrel < \over \sim \;$}
\def\simlt{\lower.5ex\hbox{\ltsima}}
\def\gtsima{$\; \buildrel > \over \sim \;$}
\def\simgt{\lower.5ex\hbox{\gtsima}}
\def\eq{equation}
\def\ion#1#2{{\rm #1\,\sc #2}}
\def\HI{{\ion{H}{i} }}
\def\HII{{\ion{H}{ii} }}

\def\GII{{\ion{He}{ii} }}

\def\hide#1{}
\def\bi{\begin{itemize}}
\def\ei{\end{itemize}}

\begin{document}

\pagestyle{myheadings} \markright{DRAFT: \today\hfill}

\title{Effect of Primordial Black Holes on the Cosmic Microwave
  Background and Cosmological Parameter Estimates}

\author{Massimo Ricotti}
\affil{Department of Astronomy, U of Maryland, College Park,
  MD 20742} 
\email{ricotti@astro.umd.edu}

\author{Jeremiah P. Ostriker and Katherine J. Mack}  
\affil{Department of Astronomy, Princeton University;}

\begin{abstract}
We investigate the effect of non-evaporating primordial black holes
(PBHs) on the ionization and thermal history of the universe.  X-rays
emitted by gas accretion onto PBHs modify the cosmic recombination
history, producing measurable effects on the spectrum and anisotropies
of the Cosmic Microwave Background (CMB).  Using the third-year WMAP
data and FIRAS data we improve existing upper limits on the
abundance of PBHs with masses $>0.1$ M$_\odot$ by several orders of
magnitude.

Fitting WMAP3 data with cosmological models that do not allow for
non-standard recombination histories, as produced by PBHs or other
early energy sources, may lead to an underestimate of the best-fit
values of the amplitude of linear density fluctuations ($\sigma_8$)
and the scalar spectral index ($n_s$).  Cosmological parameter
estimates are affected because models with PBHs allow for larger
values of the Thomson scattering optical depth, whose correlation with
other parameters may not be correctly taken into account when PBHs are
ignored. Values of $\tau_e \sim 0.2$, $n_s \sim 1$ and $\sigma_8 \sim
0.9$ are allowed at 95\% CF. This result that may relieve recent
tension between WMAP3 data and clusters data on the value of
$\sigma_8$.

PBHs may increase the primordial molecular hydrogen abundance by up to
two orders of magnitude, this promoting cooling and star
formation. The suppression of galaxy formation due to X-ray heating is
negligible for models consistent with the CMB data. Thus, the
formation rate of the first galaxies and stars would be enhanced by a
population of PBHs.
\end{abstract}
\keywords{cosmology: theory --- cosmology: observations --- early
  universe --- cosmic microwave background --- cosmological parameters
  --- black hole physics}

\section{Introduction}

During the radiation era, before the formation of the first stars
and galaxies, small perturbations of the energy-density of the
universe on scales comparable to the particle horizon may become
gravitationally unstable. The outcome of the collapse can be the direct
formation of primordial black holes (PBHs) \citep{Hawking:71, Carr:74,
Musco:05, Harada:05}.  If these PBHs form at a sufficiently high mass,
they do not evaporate but instead begin to grow by accretion, producing
x-rays.

A population of sufficiently massive and numerous PBHs may provide an
important and observable source of energy injection into the cosmic
gas before the formation of non-linear large scale structures and
galaxies. Several previous papers have addressed this scenario
\citep{Carr:81, GnedinOR:95, MillerO:01}. Here, we improve on previous
calculations by modelling the accumulation of dark matter around PBHs,
the proper motion of PBHs and feedback effects including Compton
drag. We simulate the ionization and thermal history of primordial
plasma using semi-analytical calculations as in
\cite{RicottiO:03}. The early energy injection by PBHs may produce
observable distortions of the CMB spectrum \citep{Battistelli:00} and
may affect CMB anisotropies. We fit cosmological models which include
the effect of PBHs on the cosmic recombination history to the 3rd year
WMAP data \citep{Spergel:06}.  This is done by modifying the
Monte-Carlo code COSMOMC \citep{Lewis:02} appropriately. Our main goal
is to improve existing upper limits on the mass and abundance of
non-evaporating PBHs. In addition, we find that the existence of PBHs
may affect the best fit estimates of cosmological parameters. In
particular, the value of $\tau_e$, $\sigma_8$ and $n_s$ can be
underestimated in models which do not account for the existence of
PBHs.  This result is more general than the case of PBHs discussed in
this paper. Any source of energy injection at early time that modifies
the recombination history may lead to underestimating $\tau_e$, $n_s$ and
$\sigma_8$ if the effect is excluded a priori when fitting the WMAP3
data. This scenario may ease recent tensions between the WMAP3 analysis
that favors low values of $\sigma_8 \sim 0.74$ (assuming a standard
recombination history) and clusters data that may favor larger values of
$\sigma_8 \sim 0.9$ \citep{Evrard:07}, but see also \citep{BodeO:07}.

Finally, ionization from the X-rays of accreting PBHs will increase
the amount of H$_2$ significantly and the resultant extra cooling will
enhance early star formation.

This paper is organized as follows. In \S~\ref{sec:PBH} we review the
properties of PBHs and the current constraints.
In \S~\ref{sec:eq} we introduce the
basic equations for gas and dark matter accretion and for the accretion
luminosity of PBHs. In \S~\ref{sec:feedback} we evaluate the effect of
local and global feedback processes. In \S~\ref{sec:res} we present
the results of calculations of the ionization and thermal history of
the IGM and in \S~\ref{sec:obs}, data from WMAP3 and COBE are used to
constrain the mass and abundance of PBHs. A summary and discussion is
presented in \S~\ref{sec:sum}.  Throughout the rest of the paper we use
the following cosmological parameters ($h=0.73, \Omega_m=0.238 ,
\Omega_b=0.0418 , \sigma_8=0.74, n_s=0.95$) from WMAP3
\citep{Spergel:06}.

\section{Primordial black holes}\label{sec:PBH}

In the Newtonian regime, the theory of PBH formation can be
understood in simple terms. The Jeans length in a static and
homogeneous fluid with sound speed $v_s$ is $R_J \sim (c/v_s)R_{Sch}$,
where $R_{Sch}$ is the Schwarzschild radius of a black hole with
density equal to the mean cosmic value. During the radiation era $v_s
\approx c/\sqrt{3}$, thus $R_J \rightarrow R_{Sch}$.  This means that
the self gravitating regime appears when the perturbation is very
close to the black hole regime.  The critical overdensity needed to
trigger PBH collapse is $w \simlt 1/3$, where $P=w
\rho c^2$ is the cosmic equation of state \citep{Carr:75,
Green:04}. The typical mass of a PBH is approximately equal to or smaller
than the mass within the particle horizon at the redshift of its
formation, $z_f$:
\begin{equation}
{M_{pbh} \over M_{H}(z_{eq})} \sim f_{Hor}\left({1+z_{eq} \over
1+z_{f}}\right)^{2} \sim f_{Hor}\left({\beta \over
\Omega_{pbh}}\right)^2.
\label{eq:relic}
\end{equation}
Here, $f_{Hor}=M_{pbh}/M_{H}(z_f) \le 1$ is the ratio of the mass of
the PBH to the mass of the particle horizon at $z=z_f$, $\beta$ is the
density parameter of PBHs at $z=z_f$, $M_{H}(z_{eq}) \sim 3.1 \times
10^{16}$ M$_\odot$ is the mass of the horizon at the redshift of
matter-radiation equality $z_{eq} \approx 3000$ \citep[for $\Omega_m
h^2=0.127$,][]{Spergel:06}. The relationship
$\beta(1+z_f)=\Omega_{pbh}(1+z_{eq})$, valid for non-evaporating PBHs,
relates the density parameter of PBHs today,
$\Omega_{pbh}(M)=\rho_{pbh}/\rho_{crit}$, to the one at the time of
formation, $\beta(M)$. For example, if a tiny fraction $\beta \sim
10^{-9}$ of the cosmic energy-density collapses into PBHs during the
quark-hadron phase transition at $t \sim 10^{-5}$~s, it follows from
\eq~(\ref{eq:relic}) that about 30\% of the present day dark matter is
made of PBHs with mass $M_{pbh} \sim 1$ M$_\odot$ (we assumed
$f_{Hor}=1$).  Nothing prevents the dark matter from being a mixture
of weakly interacting particles (WIMPs) and PBHs with
$\Omega_{dm}=\Omega_{pbh}+\Omega_{wimp}$. However,
\eq~(\ref{eq:relic}) shows that fine tuning of the value of $\beta$ is
required in order to have $\Omega_{pbh} \sim \Omega_{wimp}$. Thus, it
is often considered more likely that dark matter is dominated by
either WIMPs or PBHs.  PBHs may have an extended mass range if they
form at different epochs and, even if they all form at one epoch,
simulations show that their mass distribution is broad
\citep[\eg,][]{Choptuik:93, Evans:94}.  Thus, present-day dark matter
may be composed of a mixture of relatively massive and tiny PBHs, some
of which may be completely evaporated or have left Planck-mass relics.
In practice, using astrophysical tests only, sufficiently small PBHs
would be virtually indistinguishable from WIMPs.  It is worth pointing
out that no fine tuning of $\beta(z_f)$ is required to produce
$\Omega_{dm}=\Omega_{pbh} \sim 1$. Any value of $\beta \le 1$ and
$z_f$ is allowed but, of course, different values of the entropy of
the Universe and different redshift of equivalence would be
produced. Specifically, the radiation density parameter today is
related to $\beta$ and $z_f$ by the relationship
$\Omega_{rad,0}=\Omega_{dm}/(1+z_{eq}) \sim
\beta^{-1}(1+z_f)^{-1}$. In addition, a scenario in which PBHs
produced after inflation evaporate leaving only Planck mass relics may
explain why $\Omega_{rad,0}h^2 \sim 4.35 \times 10^{-5}$ and provide a
justification for the large value of the entropy of the Universe
\citep{AlexanderM:07}.

PBHs are a unique probe of the early universe, of high energy
processes and of quantum gravity. The small mass scales at which PBHs may
form are inaccessible to Cosmic Microwave Background (CMB)
experiments. As of today there is no solid evidence for the existence
of PBHs, but their presence would be very difficult to detect even if
they constitute the bulk of the dark matter. Early results from the
MACHO collaboration \citep{Alcock:00} suggested a possible detection
of solar mass objects constituting about 20\% of the Galactic halo
mass. Recent results seem to disfavor this claim
\citep[\eg,][]{Hamadache:06}. The goal of the present work is to
better constrain the abundance of relatively massive ($M \simgt 0.1$
M$_\odot$) PBHs by modeling their effect on the ionization history of
the universe.  To further motivate the present study we briefly
review the current status of theoretical works and observational
limits on the existence of PBHs. For a more comprehensive review see
\cite{Carr:05}.

\subsection{Current observational limits on PBHs}

Many physical processes may lead to the formation of PBHs.  For example,
PBHs may form from perturbations after inflation,
during phase-transitions of the cosmic equation of state or from
topological defects \citep[][\eg]{CarrH:74, Carr:05}. Theoretically,
it is unclear what is the largest allowed mass of PBHs.  Inflationary
theories predict an almost scale-invariant ($n \sim 1$) initial
spectrum of perturbations. This spectrum has the remarkable property
that all perturbations have the same amplitude when they enter the
horizon. Thus, in this case, it may appear that the probability of PBH
formation is almost independent of their mass. Small-mass PBHs would
have a larger probability of formation if the spectrum were slightly
blue (with $n>1$) and vice versa if the spectrum were red, as seems to
be indicated by available data \citep{Spergel:06, Chongchitnan:07}.
\cite{Jedamzik:97} have shown that during a first-order phase
transition the cosmic equation of state may become softer ($w \ll
1/3$). As a result, the formation of PBHs may be substantially
enhanced at that particular mass scale. PBHs with masses $\sim 1$
M$_\odot$ may form during the QCD (quark-hadron) phase transition at
$t \sim 10^{-5}$~s, or PBHs with mass $10^5$ M$_\odot$ may form during
$e^+-e^-$ annihilation era. PBHs with masses of $100-1000$ M$_\odot$
may be produced in two-stage inflationary models designed to fit the
low WMAP quadrupole \citep{Kawasaki:06}. One interesting feature of
this model is a huge bump in the power spectrum at kpc scales with
overdensity $\delta >1$. Other possible scenarios for PBH
formation involve non-Gaussian perturbations produced by cosmic
strings, cosmic string collapse or bubble collisions following
second-order phase transitions
\citep[\eg,][]{Polnarev:91,Rubin:01,Stojkovic:05, Nozari:07}. It is
generally considered unlikely that PBH formation occurred after
$t=1$~s, when $M_{pbh} \sim M_H \simgt 10^5$ M$_\odot$, because the
physics in this domain is sufficiently understood and their formation
would affect primordial nucleosynthesis.

Observationally, the abundance of PBHs is well constrained only at very
small masses. PBHs with masses smaller than $5 \times 10^{14}$ g should
not exist today because they evaporate in less than a Hubble time by
emission of Hawking radiation \citep{Hawking:74}. The radiation from
evaporating PBHs affects nucleosynthesis and the CMB spectrum and may
overproduce the observed gamma-ray background. Hence, upper limits on
the abundance of PBHs with masses $1~{\rm g} < M_{pbh} < 10^{15}$ g
are quite stringent: $\beta(M) \sim 10^{-20}-10^{-22}$
\citep[\eg][]{Carr:03}.
The number density of PBHs with masses larger than $10^{15}$ g is
poorly constrained because they emit negligible amounts of Hawking
radiation (their timescale for evaporation is longer than the age of
the Universe today). Dynamical constraints \citep{LaceyO:85, Carr:94,
More:93, Jin:05}, effects on the matter power spectrum \citep{Afshordi:03} and
statistical studies of binary stars in the Galactic halo
\citep{Yoo:04}, have been used to rule out PBHs more massive than
$1000$ M$_\odot$ as main constituent of the dark matter, but
\cite[][hereafter Paper~I]{Mack:06}, summarizing these constraints,
find that a
domain remains for which massive PBHs can make a significant
contribution to the dark matter.  Searches for microlensing events
toward the Large Magellanic Cloud by the MACHO and EROS collaborations
so far provide the most stringent constraints on the existence of
non-evaporating
PBHs. They have ruled out PBHs as the bulk of the galactic dark matter
in the mass range $10^{-7}~{\rm M}_\odot<M<30$ M$_\odot$
\citep{Alcock:98, Alcock:01}. \cite{Alcock:00} have claimed a positive
detection of Massive Compact Halo Objects (MACHOs) with $M \sim 0.1-1$
M$_\odot$ constituting about $20\%$ of the Milky-Way dark matter
halo. More recent works have not confirmed this result
\citep{Hamadache:06} but this mass range is of particular interest
because it is about the mass of PBHs that may form during quark-hadron
phase transition as proposed by \citep{Jedamzik:97}. The gray curves
in Figure~\ref{fig:FIRAS}(left) summarize the current status of
observational upper limits on the abundance of PBHs with
$M_{pbh}>10^{15}$ g. The two black curves labeled ``FIRAS'' and
``WMAP3'' show the new upper limits derived in the present work.

\subsection{PBH growth and ``clothing'' dark halo}

PBHs more massive than $10^5$ M$_\odot$ may only exist if substantial
accretion takes place after their formation or if many PBHs which form
in a cluster merge into a larger one
\citep[][]{Meszaros:75a,Khlopov:05, Chisholm:06}. In the Newtonian
approximation, PBHs can only grow if their mass is comparable to the
horizon mass. In this case their mass increases at the same rate as
the horizon mass \citep{Zeldovich:67}. Relativistic calculations by
\cite{CarrH:74} have shown that growth by accretion is always
slower than the growth of the horizon mass. Thus, it is generally
accepted that the mass of PBHs does not increase significantly after
their formation. 
In models in which the equation of state becomes stiff $w \sim 1$
\citep{Lin:76} or in brane world cosmologies in which extra dimensions
are sufficiently large, growth by accretion could be significant and
in principle may lead to the production of massive PBHs starting from
smaller seeds \citep{Bicknell:78, Guedens:02a, Guedens:02b,
Majumdar:03, Tikhomirov:05}.

Scenarios in which the bulk of the dark matter is not made of PBHs are
particularly interesting because, after the redshift of
matter-radiation equivalence, each PBH seeds the growth of a dark halo
that, over time, becomes considerably more massive than the PBH in its
center (Paper~I). The halo gravitational potential may increase the
gas accretion rate onto the central PBH by several orders of
magnitude. If the accretion is near the Eddington rate, the radiation
efficiency of PBHs may be large. At later times, when PBHs are
accreted by the massive halos of galaxies, they likely lose much of
their dark matter ``clothing'' and become ``naked'' again.

\section{Basic Equations}\label{sec:eq}

\begin{figure*}[t]
\epsscale{1.1}
\plottwo{\figname{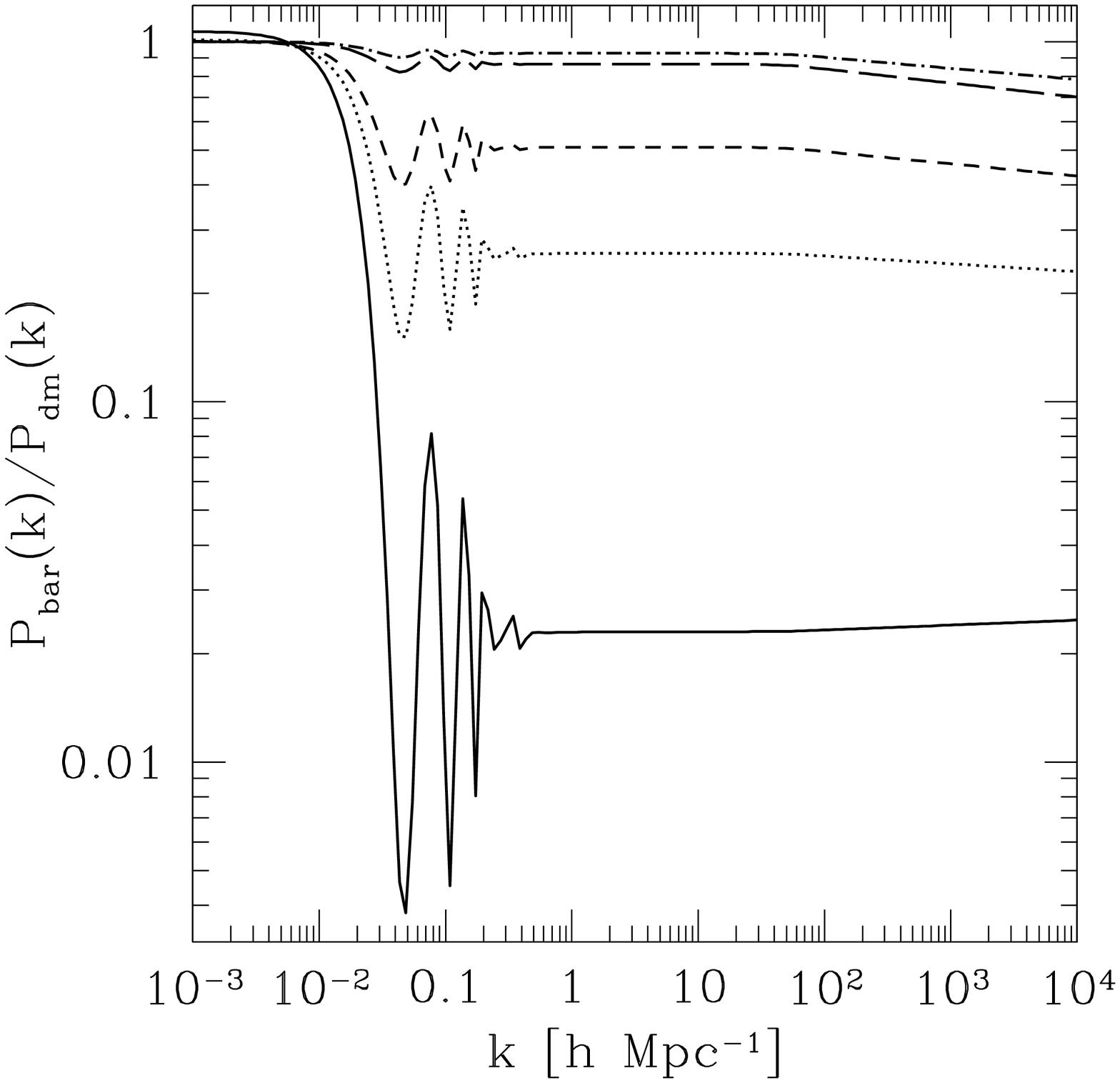}}{\figname{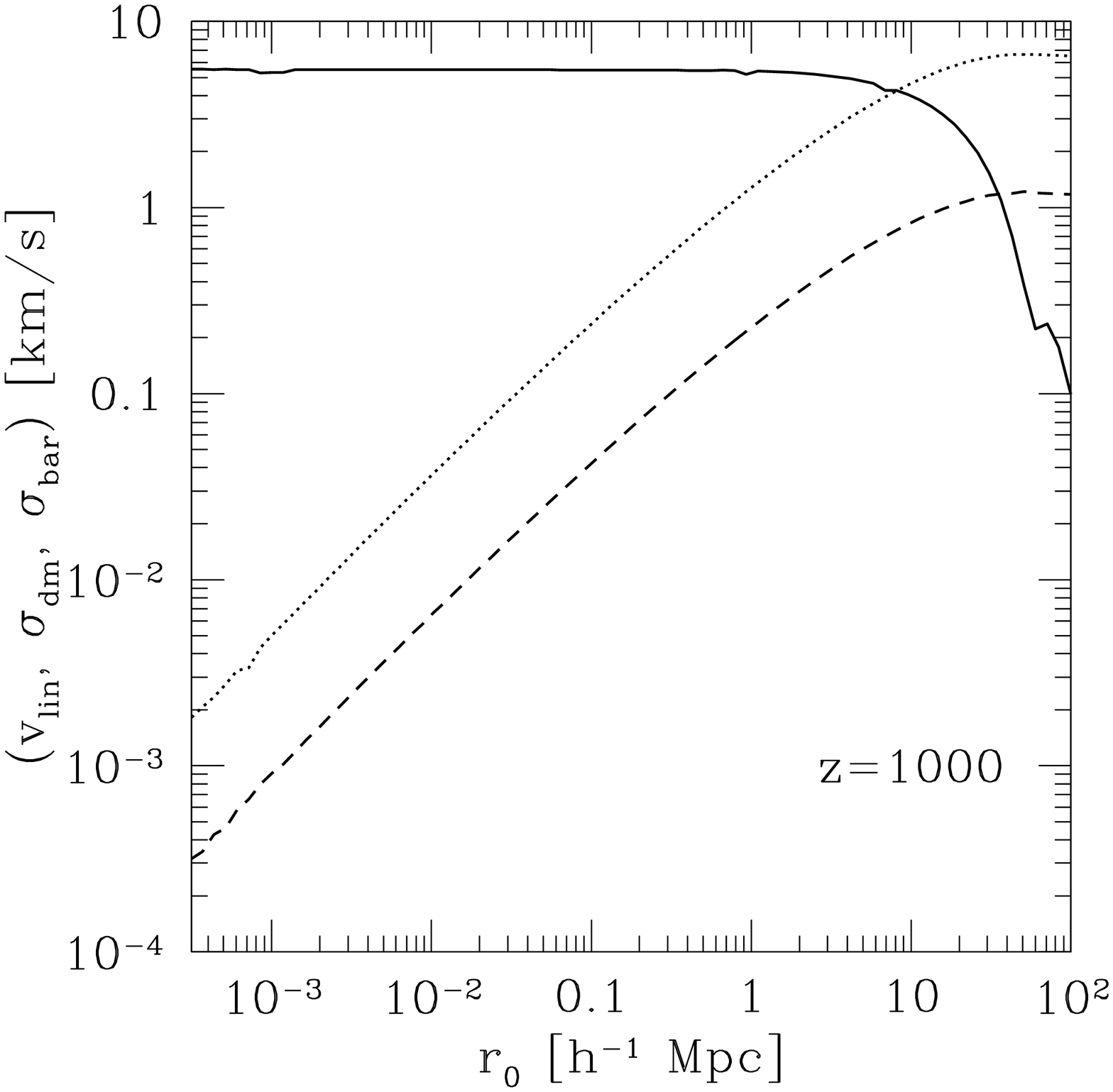}}
\caption{{\it (Left).} Ratio of the baryonic to dark matter power
  spectrum as a funtion of wavenumber $k$. Each curve, from bottom to
  the top, corresponds to scale factors $a=0.001, 0.002, 0.005, 0.01,
  0.05, 0.1$, respectively. {\it (Right).} The solid curve shows the
  mean relative velocity between the dark matter and baryons in a
  sphere of comoving radius $r_0$. The dashed and dotted curves show
  the velocity dispersion within a sphere of comoving radius $r_0$ for
  the baryons and for the dark matter, respectively.}
\label{fig:power}
\end{figure*}

Here we review the theoretical framework for the accretion of dark
matter and baryons by primordial black holes.
A point mass, $M$, immersed in an hydrogen gas with constant number
density $n_{gas}$ and sound speed $c_{\rm s}$ that moves at velocity
$v$ with respect to the gas accretes at a rate
\[
\dot M_{\rm b}=\lambda 4 \pi m_H n_{gas}
v_{eff} r_B^2,
\]
where $r_B \equiv GM v_{eff}^{-2}$ is the Bondi-Hoyle radius and $v_{eff}
\equiv (v^2+c_s^2)^{1/2}$ \citep{BondiH:44}. The numerical values of the
Bondi radius and accretion rate are:
\begin{eqnarray}
r_B &\approx& 1.3 \times 10^{-4}
~{\rm pc}\left({M \over 1~M_\odot}\right) \left({
v_{eff} \over {\rm 5.7 km~s^{-1}}}\right)^{-2},\\
\dot M_{\rm b} &\approx&
2 \times 10^{12}~{\rm g~s^{-1}} \lambda n_{gas} \left({M \over 1~
  M_\odot}\right)^2 \left({v_{eff} \over {\rm 5.7 km~s^{-1}}}\right)^{-3}.
\label{eq:bondi}
\end{eqnarray}
The mean cosmic gas density is
\begin{equation}
n_{\rm gas} \simeq 200 ~{\rm cm}^{-3}\left({1+z \over 1000}\right)^3.
\end{equation} 
Before decoupling the gas temperature is roughly equal to the
temperature of the CMB.  Afterwards the gas temperature decreases
adiabatically due to the Hubble expansion of the Universe.  The
temperature of the IGM (neglecting for the moment the feedback
produced by accreting PBHs) is well approximated by the
fitting formula
\begin{equation}
T_{\rm gas}=(2730 ~K) \left({z+1\over 1000}\right){a_{dec} \over
(a^\beta+a_{dec}^\beta)^{1/\beta}},
\label{eq:temp}
\end{equation} 
where $a \equiv 1/(1+z)$ is the scale parameter, $a_{dec}$ is the scale
parameter at decoupling where $z_{dec} \simeq
132(\Omega_{b}h^2/0.022)^{2/5}$ and $\beta=1.72$. The gas sound speed
is $c_s=(5.7~{\rm km~s^{-1}}) (T_{gas}/2730)^{1/2}$. Thus, from
\eq~(\ref{eq:temp}) we have
\begin{equation}
c_s \approx (5.7~{\rm km~s^{-1}}) \left({1+z \over
  1000}\right)^{1/2} ~{\rm for} z \gg
z_{dec} \sim 132.
\label{eq:cs}
\end{equation}
For spherical accretion onto a point mass, assuming a non-viscous
fluid, the parameter $\lambda$ is of order unity. In the present study,
we need to take into account the effects of the growth of a dark halo
around PBHs, the Hubble expansion and the coupling of the CMB
radiation to the gas through Compton scattering (Compton drag). All
these effects can be folded into the calculation of the accretion
eigenvalue $\lambda$.  In a companion paper, \citep[][hereafter
Paper~II]{Ricotti:06}, we have derived analytical relationships for
$\lambda$ for the cases of a point mass and an extended dark halo,
including the aforementioned cosmological effects. In the following
calculations we will use the results from Paper~II to derive the
accretion rate onto ``naked'' PBHs (appropriate only if PBHs
constitute the bulk of the dark matter) and for the general case of
accretion onto PBHs including the growth of a ``clothing'' dark halo
of mass $M_h$.

To obtain this result, we must first estimate the dark matter accretion
(\S \ref{ssec:dm}) and the effects of PBH velocity
(\S \ref{ssec:propermotion}), the angular momentum (\S \ref{ssec:ang})
of the dark and baryonic matter and PBH clustering
(\S \ref{ssec:clustering}).  From the gas accretion rate, we
can then estimate the accretion luminosity, which will ultimately be
responsible for altering the evolution of the IGM.

\subsection{Growth of a dark halo around PBHs}\label{ssec:dm}

If PBHs do not constitute the bulk of the dark matter, they seed the
accumulation of a dark halo of WIMPs which grows proportionally to
$t^{2/3}$ \citep{Bertschinger:85, Mack:06}. A similar result holds if
the dark matter is made only of PBHs but with a very broad
distribution of masses.

After the redshift of equivalence the mass of the dark halo
surrounding PBHs grow proportionally to the cosmic scale parameter:
\begin{equation}
M_{h}(z)=\phi_i M_{pbh}\left(z+1 \over 1000\right)^{-1},
\end{equation}
where $\phi_i =3$ (see Paper~I). The growth of the PBH during the
radiation era is of order of unity. After $z \sim 30$, depending on
the environment, the dark halo may stop growing: if a PBH evolves in
isolation it can continue to grow, otherwise it will lose mass due to
tidal forces as it is incorporated into a larger galactic dark halo.

If $M_{h} > M_{pbh}$ and the accretion is not perfectly spherical, the
dark halo has a self-similar power-law density profile $\rho \propto
r^{\alpha}$ with $\alpha \sim -2.25$ \citep{Bertschinger:85} truncated
at a halo radius
\begin{equation}
r_{h} = 0.019~{\rm pc}\left({M_{h} \over
1~M_\odot}\right)^{1/3}\left({1+z \over 1000}\right)^{-1},
\label{eq:rhalo}
\end{equation}
where $r_h$ is about one third of the turn-around radius (Paper~I and II).  

\subsection{Proper motions of PBHs}\label{ssec:propermotion}

The motion of the accreting mass relative to the surrounding material
strongly affects the accretion rate (see equation [\ref{eq:bondi}]).
This proper motion is determined by the relative amplitude of
inhomogeneities of dark matter and gas.  More precisely, assuming that
PBHs behave like dark matter particles and neglecting binary
interactions, we estimate their relative velocity with respect to the
gas in the linear and non-linear regimes. We show that at redshifts $z
\simgt 30-50$, due to the imperfect coupling of gas and dark matter
perturbations (\ie, Silk damping), the peculiar velocity of PBHs with
respect of the gas is of the order of the gas sound speed. At
redshifts $z<30$ the growth of non-linear perturbations dominates the
motion of PBHs. The velocity of PBHs falling into the potential wells
of the first galaxies is sufficiently large to virtually stop the
accretion of gas from the intergalactic medium onto PBHs, until they
came to rest at the center of the halo into which they are accreted.

\subsubsection{Linear regime}\label{ssec:v_lin}

In linear theory, gas and dark matter perturbations are coupled. Before
the redshift of ``decoupling'' at $z \sim 100$, the growth of gas
inhomogeneities on small scales is suppressed by Silk damping
\citep{Silk:68}. At these scales and redshifts, the gas flow lags
behind the dark matter until it is able to catch up at later
times. This process is illustrated quantitatively in the left panel of
\fig~\ref{fig:power}, which shows the ratio of the gas to dark matter
power spectra in our fiducial $\Lambda$CDM model.  The curves, from
bottom to the top, show the ratio of the power spectra of the baryons
to dark matter calculated using the code ``lingers'' in the Graphic1
package \citep{Bertschinger:85} at scale parameters $a=0.001, 0.005,
0.01, 0.05$, and $0.1$, respectively.

>From the power spectra of dark matter and baryons we calculate the
ensemble average of the center-of-mass velocity of a patch
of the universe of comoving radius $r_0$ \citep{OstrikerS:90}:
\begin{equation}
\langle V_{i}\rangle^2 = {\Omega_m^{1.2} H^2 \over 2\pi^2}\int_0^\infty
P_{i}(k) w_s^2(k,a)w_l^2(k,r_0) dk,
\label{eq:vmean}
\end{equation}
where $\Omega_m$ is the cosmological density parameter, $w_s$ and
$w_l$ are window functions (here we use ``top hat'' window functions)
and $a$ is a small scale smoothing of the perturbations. The choice
of the value of $a$ is not critical as long as $a \ll r_0$. The index
$i=dm, bm$ refers to dark matter and baryons, respectively. The
ensemble average of the velocity variance within a patch of comoving
radius $r_0$ is calculated in a similar fashion:
\begin{equation}
\langle \sigma_{i}\rangle^2 = {\Omega_m^{1.2} H^2 \over 2\pi^2}\int_0^\infty
P_{i}(k) w_s^2(k,a)[1-w_l^2(k,r_0)] dk.
\label{eq:sigma}
\end{equation}
This equation will be used later (\S~\ref{ssec:ang}) to estimate
the angular momentum of gas and dark matter accreting onto PBHs. The
``cosmic Mach number,'' ${\cal M}_i=\langle V_i\rangle/\langle
\sigma_i\rangle$, is independent of the uncertainties on the
normalization of the power spectrum \citep{OstrikerS:90}.

Since the flow of the gas and the dark matter trace each other, their
mean relative velocity is $\langle V_{rel}\rangle \equiv \langle
V_{dm}\rangle-\langle V_{bar}\rangle$. The right panel of
\fig~\ref{fig:power} shows the mean relative velocity $\langle
V_{rel}\rangle$ of PBHs with respect to the baryons (solid curves),
and the velocity dispersion $\langle \sigma_{dm,bm}\rangle$ for the
gas (dashed curve) and the dark matter (dotted curve) within a sphere
of comoving radius $r_0$ at $z=1000$.  Power-law fits for $\langle
\sigma\rangle$ as a function of $r_0$ and $z$ are:
\begin{eqnarray}
\sigma_{bm}&\approx& (0.35~{\rm km~s}^{-1})~\left({r_0\over 1~{\rm Mpc}}\right)^{0.85}\left({1+z \over
1000}\right)^{-1},\\
\sigma_{dm}&\approx& (1.58~{\rm km~s}^{-1})~\left({r_0\over 1~{\rm Mpc}}\right)^{0.85}\left({1+z \over
1000}\right)^{-{1\over 2}}.
\label{eq:sig_fits}
\end{eqnarray}
The fits are accurate to about 5\% for $r_0<1$ Mpc in the redshift range
$50<z<2000$.

\begin{figure}[t]
\epsscale{1.2}
\plotone{\figname{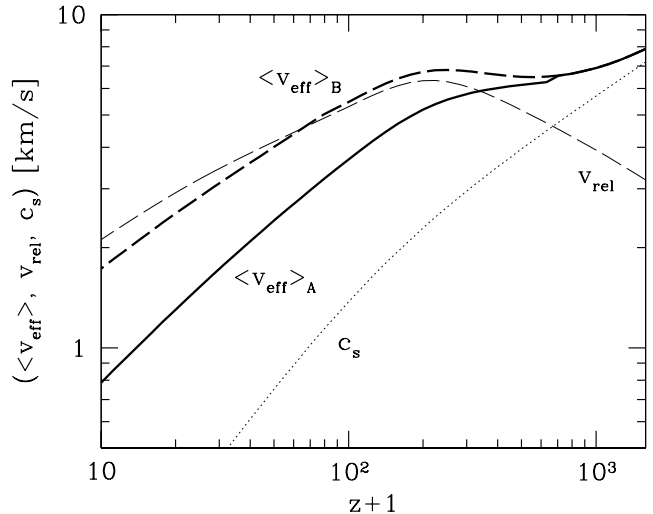}}
\caption{Luminosity weighted effective velocity of PBHs (thick
  curves): $\langle v_{eff}\rangle_A$ (solid curve), is weighted
  assuming $l \propto \dot m^2$ and $\langle v_{eff}\rangle_B$ (dashed
  curve), is weighted assuming $l \propto \dot m$. See the text for
  details. The thin curves show the variance of the velocity
  distribution, $\langle V_{rel}\rangle$ (dashed line) and the gas
  sound speed, $c_s$ (dotted line), respectively.}
\label{fig:vel}
\end{figure}

In \S~\ref{ssec:bol} we discuss in more detail the radiative
efficiency of accreting PBHs. We anticipate that the radiative
efficiency $\epsilon$ and the accretion luminosity $L$ depend on the
accretion rate $\dot m$: $L \propto v_{eff}^{-6}$ if $\dot m<1$ and
$L \propto v_{eff}^{-3}$ if $\dot m \simgt 1$, where the effective
velocity $v_{eff}$ is defined in \S~\ref{sec:eq} (see
\eq~[\ref{eq:bondi}]).  The effect of the linear velocity field, which
is Gaussian, on the mean accretion luminosity of PBH is given by the
statistical ensemble average of $v_{eff}^{-\alpha}$ weighted by the
distribution function of the PBHs relative velocities with respect to
the gas:
\begin{equation}
\langle v_{eff}^{-\alpha} \rangle= \int_0^{\infty}{f_M(v,\sigma) dv
  \over (c_s^2 + v^2)^{\alpha/2}},
\label{eq:int1}
\end{equation}
where $\alpha=3$ or $\alpha=6$, depending on the value of the
accretion rate. The linear velocity field is Gaussian; thus the
distribution of the moduli of the 3D velocity field is a Maxwellian,
$f_M(v, \sigma)$ with $\sigma=\langle V_{rel}\rangle$. Finally,
integrating \eq~(\ref{eq:int1}) and defining ${\cal M}_{pbh}=\langle
V_{rel}\rangle /c_s$, we find
\begin{equation}
\langle v_{eff} \rangle_A \approx
\cases{
\langle c_s \rangle \left[{{16 \over \sqrt{2\pi}}{\cal
      M}_{pbh}^{3}}\right]^{1 \over 6} & {for ${\cal M}_{pbh}>1$,}\cr 
c_s(1+{\cal M}_{pbh}^2)^{1 \over 2} & {for ${\cal M}_{pbh}<1$,}\cr
}
\label{eq:alpha_6}
\end{equation}
for $\alpha=6$ and
\begin{equation}
\langle v_{eff} \rangle_B \approx
\cases{
c_s{\cal M}_{pbh}\left[\sqrt{{2 \over \pi}} \ln({2\over e}{\cal M}_{pbh})\right]^{-{1\over 3}} &
  {for ${\cal M}_{pbh}>1$},\cr 
c_s(1+{\cal M}_{pbh}^2)^{1/2} & {for ${\cal M}_{pbh}<1$,}\cr
}
\label{eq:alpha_3}
\end{equation}
for $\alpha=3$. The calculations presented in this section on the
effect of linear perturbations on the accretion luminosity of PBHs are
summarized in \fig~\ref{fig:vel}.  The thick solid and dashed curves
show $\langle v_{eff}\rangle_B$ from \eq~(\ref{eq:alpha_3}), and
$\langle v_{eff}\rangle_A$ from \eq~(\ref{eq:alpha_6}) respectively,
as a function of redshift. For comparison, the thin dashed and dotted
curves show $\langle V_{rel} \rangle$ and the sound speed of the gas,
respectively. Neglecting feedback effects, the PBH velocity is
comparable to the gas sound speed at redshifts $z \simlt 500$. On
average, the motion of PBHs in the linear regime reduces the
gas accretion rate by a factor of a few with respect to the static case.

\subsubsection{Non-linear regime}

The velocity of PBHs falling into the gravitational potential of
nearby galactic halos can be roughly estimated assuming that it is
comparable to the circular velocity of virialized halos of $2\sigma$
density perturbations: $v_p \sim v_c(2\sigma, z)$. We calculate
$v_c(2\sigma, z)$ using the Press-Schechter formalism
\citep{Press:74}.  Adopting a top-hat window function to calculate the
variance of density perturbations, we find
$\sigma(M) \approx 10.2-0.79\log(M)$ for masses
$M<10^{10}$ M$_\odot$. Hence, for the mass range of interest,
\begin{equation}
M_{2\sigma}=(8.8 \times 10^{12} ~{\rm M}_\odot)
\exp[-1.8(z+1)],
\label{eq:sig_nl}
\end{equation}
is a sufficiently accurate approximation for the mass of $2\sigma$
perturbations as a function of redshift. From \eq~(\ref{eq:sig_nl}) we
calculate the circular velocity and thus typical proper velocity of
PBHs induced by non-linear structures, as a function of redshift:
\begin{equation}
v_p \sim v_c=(17 ~{\rm km~s^{-1}}) \left({M_{2\sigma} \over 10^8
    M_\odot}\right)^{1\over3}\left({z+1 \over 10}\right)^{1\over2}.
\label{eq:v}
\end{equation}
The fraction of the dark matter and hence of PBHs in virialized halos
with mass $>M_{min}$ is $f_{vir}(z)=1-{\rm erf}(\nu_{min}/\sqrt{2})$,
where ${\rm erf}$ is the error function and
$\nu=\delta_c/\sigma(M_{min},z)$. For a fraction $f_{vir}$ of PBHs we
assume that they have a velocity $v_p$ relative to the gas and that
the gas has overdensity $\delta=200$. The remaining PBHs are in the
intergalactic medium and are subject to global and local thermal
feedback (see \S~\ref{sec:feedback}).

Dynamical friction may allow the most massive PBHs to spiral to the
center of the halos in less than a Hubble time (see \S~\ref{sec:sum}),
where they meet favorable conditions for accretion. For smaller PBHs
inside virialized halos, their contribution to the accretion
luminosity is important only in the redshift range $10\simlt z \simlt
30$.  At $z>30$ the fraction of PBHs in virialized halos is small. At
$z<10$, the first massive halos form and the rate of gas accretion
onto PBHs decreases rapidly due to the increasing relative velocity of
the PBHs with respect to the gas: $v_c \simgt 10~{\rm km s^{-1}}$.
During this redshift interval only a fraction $\sim 10-20\%$ of PBHs
is inside virialized halos. Most PBHs at $z \sim 10$ are still in the
intergalactic medium. The gas accretion onto the PBHs in the
intergalactic medium is strongly suppressed at $z<10$ due to global
thermal feedback (see \fig~\ref{fig:rei1}).

\subsection{Angular momentum of accreted material}\label{ssec:ang}

The angular momentum of the accreting gas determines whether a disk
forms or the accretion is quasi-spherical. In turn, the geometry of gas
accretion determines the radiative efficiency $\epsilon(\dot m)$, \ie,
the efficiency of conversion of gravitational energy into radiation
(see \S~\ref{ssec:bol}).  The angular momentum of accreting dark
matter determines the density profile of the dark halo enveloping a PBH
and whether the mass of a PBH can grow substantially by accreting a
fraction of its enveloping dark halo. The angular momentum of the gas
and dark matter accreting onto PBHs can be estimated from
\eq~(\ref{eq:sig_fits}) for the mean values of the velocity dispersion
within a comoving volume of radius $r_0$ (see right panel of
\fig~\ref{fig:power}).  For the gas we estimate
$\sigma_{bm}(r_{B,com})$ at the Bondi radius (see
\eq~[\ref{eq:bondi}]), and for the dark matter we estimate
$\sigma_{dm}(r_{h,com})$ at the the turn-around radius (see
\eq~[\ref{eq:rhalo}]).  If we neglect the proper motion of PBHs, the
Bondi radius of a ``naked'' PBH is approximately constant at redshifts
$z<100$: $r_{B,com} \sim (1.3 \times 10^{-7}~{\rm Mpc})(M_{pbh}/1
M_\odot)$. The proper motion of PBHs produces a small reduction of the
Bondi radius that we parameterize with the function $\xi(z)=\max[{\cal
M}_{pbh}(z),1]$, where ${\cal M}_{pbh}=\langle v_{eff}\rangle /c_s$,
is the PBH Mach number. Replacing $M_h$ with the mass of the PBH in the
previous expression, we obtain an estimate of the Bondi radius
including the enveloping dark halo.  Finally, from
\eq~(\ref{eq:sig_fits}) we obtain
\begin{eqnarray}
\sigma_{bm}&\approx& \sigma_{bm,0}\xi(z)^{-1.7}
\left({1+z \over 1000} \right)^{-1}\left({M_h
\over1~M_\odot}\right)^{0.85},\\ 
\sigma_{dm}&\approx & \sigma_{dm,0} \left({1+z \over 1000}
\right)^{-{1\over2}}\left({M_h \over1~M_\odot}\right)^{0.28},
\end{eqnarray}
where $\sigma_{bm,0}=3.8 \times 10^{-7}~{\rm km~s}^{-1}$ and
 $\sigma_{dm,0}=1.4 \times10^{-4}~{\rm km~s}^{-1}$.
\begin{figure*}[t]
\epsscale{1.1}
\plottwo{\figname{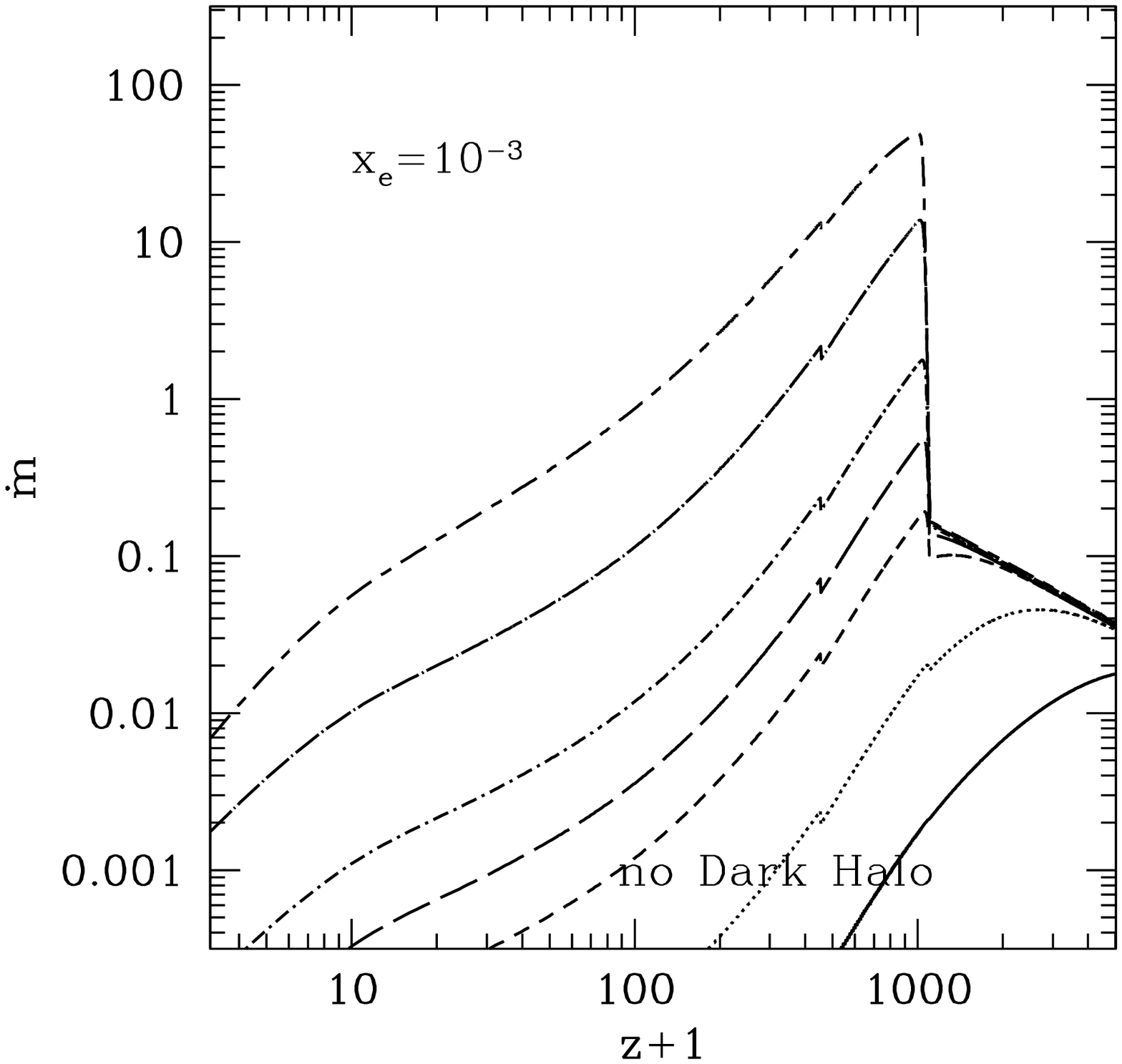}}{\figname{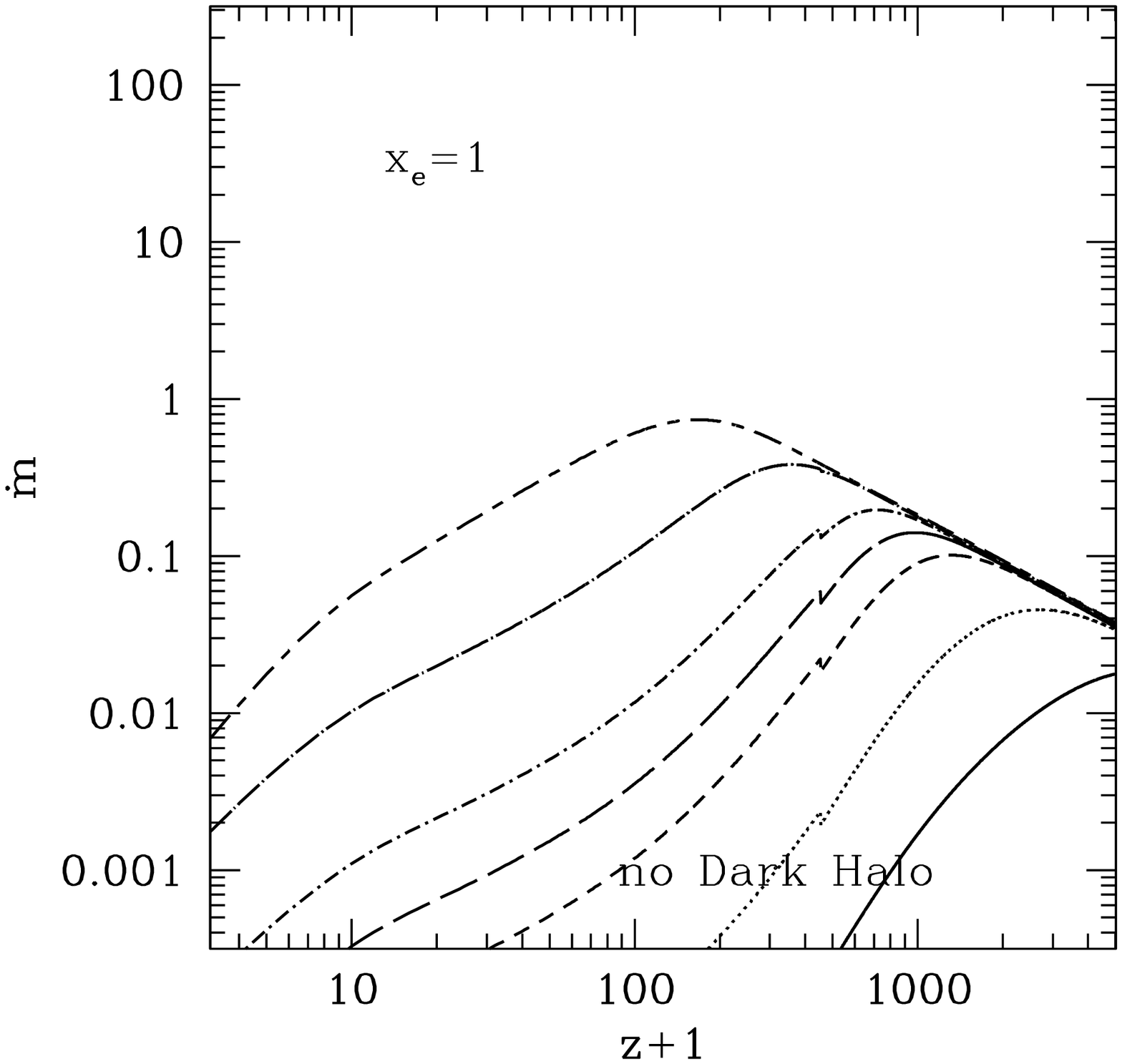}}
\caption{The dimensionless accretion rate of baryonic matter onto a
  ``naked'' PBH (without enveloping dark halo) as a function of
  redshift. The curves from bottom to top refers to $M_{pbh}=1, 10,
  100, 300, 10^3, 10^4, 10^5$ M$_\odot$. The {\it left} panel is for a gas
  with electron fraction $x_e=10^{-3}$; the {\it right} panel
  for$x_e=1$. Here, thermal feedback and the contribution from PBHs inside
  virialized halos are neglected. The motion of the PBH due to
  linear density perturbations is included.}
\label{fig:A}
\end{figure*}

Let's first consider the accretion geometry of the gas. Applying
conservation of angular momentum we find that the rotational (\ie,
tangential) velocity of the gas at a distance $r$ from the black hole
is $v(r)r \sim \sigma_{bm} r_B$, where $r_B$ is the Bondi radius.
Neglecting relativistic effects, if the velocity is much smaller than
the Keplerian velocity in the proximity of the black hole, then the
accretion is quasi-spherical; if vice versa, a disk can form.  The
Keplerian velocity at radius $r$ is $v_{kep}=c (R_{Sch}/r)^{1/2}< c$,
where $R_{Sch}=2GM/c^2$ is the Schwarschild radius of the PBH ($R_{Sch}
\sim 3~{\rm km}$ for a 1~M$_\odot$ black hole). Thus, the gas accretion
is quasi-spherical if $\sigma_{bm} < 2 D \xi(z)^2 c_s^2/c$.  We
have assumed $c_s \propto (1+z)^{1/2}$ and the constant $D \sim
1-10$ takes into account relativistic corrections.  We conclude that
the angular momentum of the gas is negligible when the halo mass is
\begin{equation}
M_h < (1746 M_\odot) D^{1.17}\xi(z)^{4.33} \left({1+z \over 1000} \right)^{2.35}.
\end{equation}
Using ${\cal M}_{pbh}(z)$ calculated in \S~\ref{ssec:v_lin} and $D=10$,
the flow is quasi-spherical for any halo mass $M_h \simlt
(1700-17000)$~M$_\odot$, almost independently of redshift. At redshifts
$z<100$, the dark halo mass is about 30 times the mass of the PBH in
its center. We conclude that an accretion disk forms at redshifts
$z<100$ around PBHs more massive than $M_{pbh} \sim
500$~M$_\odot$. Thermal feedback effects, which are important at $z <
100$, will increase our simple estimate of the critical mass for disk
formation.  So far we did not take into account feedback effects and
Compton drag which are important in reducing the gas angular
momentum. Previous studies have found that due to Compton drag the
flow is nearly spherical at redshifts $z > 100$
\citep[\eg,][]{Loeb:93,Umemura:93}.

Similar arguments apply to the dark matter component. The flow is
sufficiently spherical for the dark matter to be directly accreted
onto the PBH if $\sigma_{dm} < c D R_{Sch}/r_h$. This inequality
is satisfied for PBHs with masses
\begin{eqnarray}
M_{pbh}&>& (7.18 \times 10^5 M_\odot) D^{-2.6} \left({1+z \over 1000}
\right)^{-2.9}, ~{\rm or}\nonumber\\
{M_{pbh} \over  M_{H}}&>& (7.76 \times 10^{-13})D^{-2.6} \left({1+z \over 1000}
\right)^{-0.9},
\end{eqnarray}
in terms of the Horizon mass $M_{H}(z)$.  Including relativistic
corrections ($D \sim 10$), only PBHs with mass $\simgt 10^3$ M$_\odot$
at $z \sim 1000$ (note the steep dependence of the critical mass on
redshift) may grow substantially from direct accretion of dark matter.
The geometry of the dark halo becomes more spherical with the increasing
mass of the PBH, since the ratio of the halo radius to
the Schwarzschild radius, $r_h/R_{sw} \propto M_{pbh}^{-2/3}$,
decreases with increasing mass.  We conclude that direct accretion of
dark matter into PBHs is typically negligible. In most cases
the density profile of the dark halo is well described by the
Bertschinger self-similar solution with log slope $\alpha=2.25$.

\subsection{PBHs Clustering}\label{ssec:clustering}

Recently it has been suggested by several authors
\citep[\eg,][]{Dokuchaev:04, Carr:05, Chisholm:06} that PBHs have high
probability of forming clusters or binaries.  A precise calculation of
the effect of clustering is complex and beyond the scope of the
present paper.  Nevertheless, there are simple arguments that we can
use to show that clustering may increase the radiative efficiency
of PBHs.

We consider two regimes determined by the relationship between the
typical size of the cluster (or the the binary separation), $r_{cl}$,
and the Bondi radius of the whole system, $r_{B, tot}$.

\noindent
1) If $r_{cl}>r_{B, tot}$, the orbital velocities of the PBHs are
smaller than the effective translational velocity of the
system. Hence, we can neglect the orbital velocities to estimate the
accretion rate. To a first approximation in this case, we can ignore
the fact that the PBHs are clustered or in binary systems for the
purpose of calculating their accretion luminosity.

\noindent
2) If $r_{cl}<r_{B, tot}$, the orbital velocities of the PBHs are
larger than the effective translational velocity of the system. Thus,
the material is not accreted directly onto the PBHs but near the
center of mass of the system and subsequently onto the PBHs. Because
the PBHs are orbiting the center of mass of the cluster, the angular
momentum of the accreted material is large and so the accretion geometry is
disk-like rather then spherical. The formation of a disk can increase the
radiative efficiency by a factor of ten (see \S \ref{ssec:bol}).
Further work is
needed to better understand the accretion rate of PBHs in this regime.

\subsection{Gas accretion rate}
\begin{figure*}[t]
\epsscale{1.1}
\plottwo{\figname{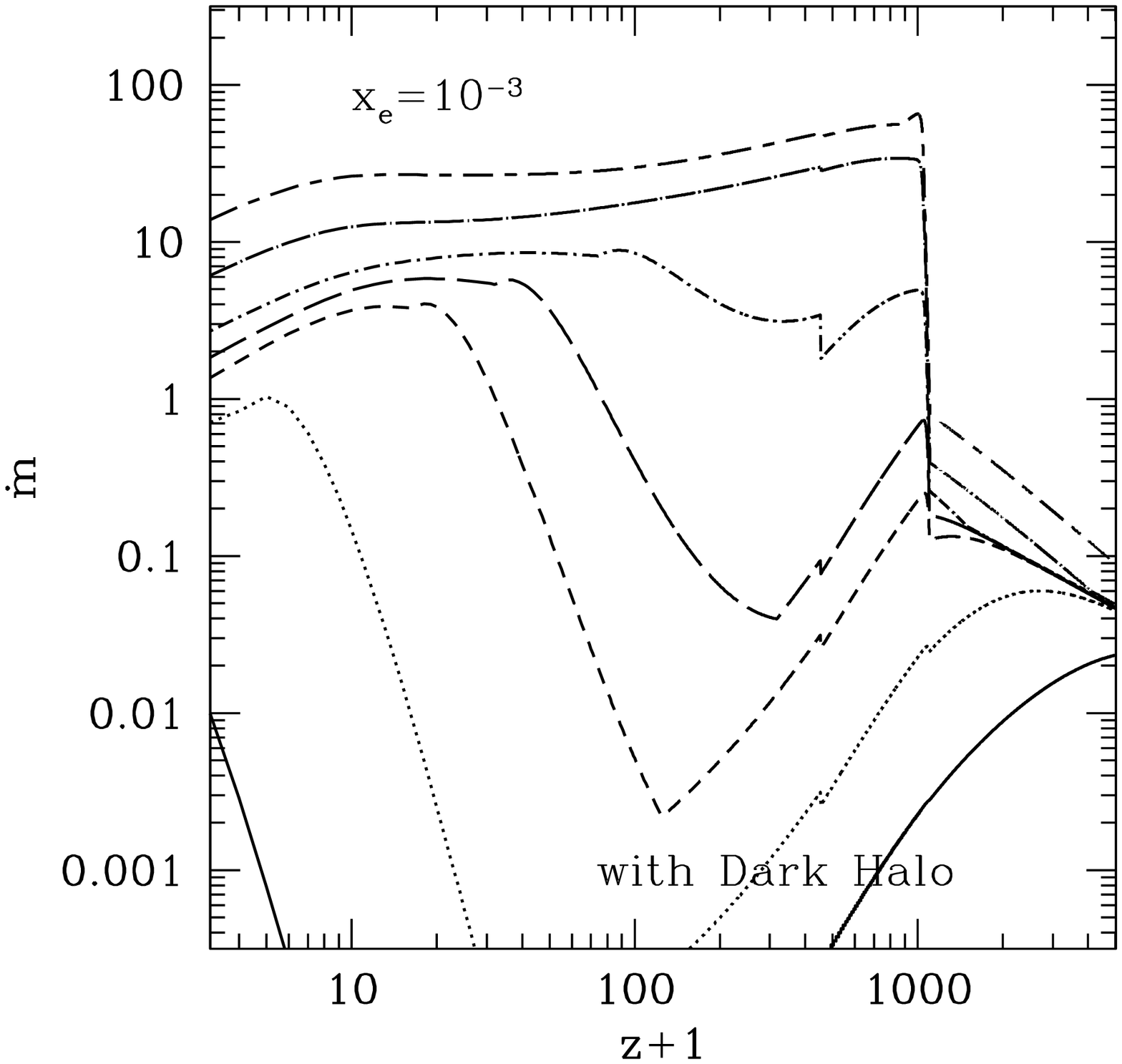}}{\figname{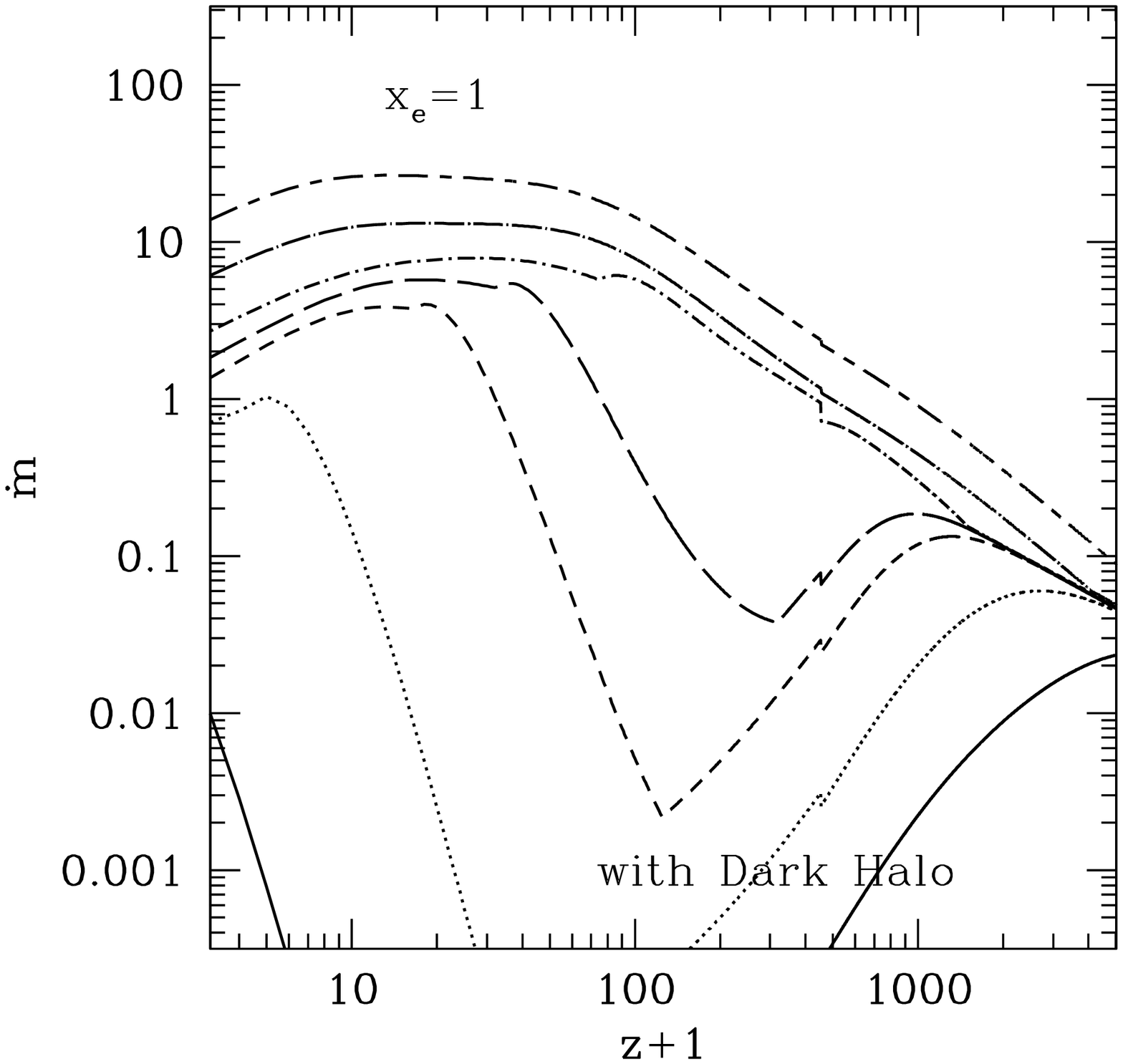}}
\caption{Same as in \fig~\ref{fig:A} but including the growth of the
  dark halo surrounding the PBH ($\alpha=2.25$ and $\phi_i=3$). The
  curves from bottom to top refers to $M_{pbh}=1, 10, 100, 300, 10^3,
  10^4, 10^5$ M$_\odot$.}
\label{fig:B}
\end{figure*}

We now have all that is needed to estimate the accretion rate onto PBHs
from \eq~(\ref{eq:bondi}). We define the dimensionless accretion
luminosity $l \equiv L_{bol}/L_{Ed}$, where $L_{Ed} \equiv 1.3 \times 10^{38}
(M_{pbh}/1~M_\odot)$ erg s$^{-1}$.  The radiative efficiency
$\epsilon$ determines the accretion luminosity for a given accretion
rate: $L_{bol}=\epsilon \dot M_{b} c^2$. Thus, defining the Eddington
accretion rate as $\dot M_{Ed} \equiv L_{Ed}/c^2=1.44 \times
10^{17}(M_{pbh}/M_\odot)$ g s$^{-1}$ and the dimensionless accretion rate
as ${\dot m} \equiv {\dot M_{b} /\dot M_{Ed}}$ we have $l=\epsilon \dot m$.

The dimensionless Bondi-Hoyle accretion rate for a PBH without a dark halo is:
\begin{equation}
{\dot m} = (1.8 \times 10^{-3} \lambda) \left({1+z \over
    1000}\right)^3 \left({M_{pbh} \over 1
    M_\odot}\right)\left({v_{eff} \over 5.74~{\rm km
    ~s}^{-1}}\right)^{-3}.
\label{eq:acc1}
\end{equation}
This equation is valid only if PBH constitute all the dark matter:
$f_{pbh}=\Omega_{pbh}/\Omega_{dm} =1$.  If the accreting object is a
point mass (\ie, a ``naked'' PBH), $\lambda$ depends only on the value
of the dimensionless gas viscosity $\hat \beta=\beta^{eff}r_B/c_{s}$:
if $\hat \beta \ll 1$ we find $\lambda \sim 1$ and if $\hat \beta \gg
1$ we have $\lambda \sim 1/\hat \beta$.  The fit to the accretion
eigenvalue for an isothermal gas is given by the formula (see Paper~II),
\begin{equation}
\lambda =\exp\left({9/2 \over 3+\hat{\beta}^{~0.75}}\right)x_{\rm
cr}^2
\label{eq:lambda1}
\end{equation}
where the dimensionless sonic radius ($x_{cr} \equiv r_{rc}/r_B$) is $x_{cr}=
[-1+(1+\hat \beta)^{1/2}]/\hat \beta$.  The effective gas viscosity,
$\beta^{eff}=\beta+H$, is the sum of two terms. The first term, due to
Compton drag, is $\beta(z)=2.06 \times 10^{-23}x_e(1+z)^4$ s$^{-1}$, and
the second, due to the cosmic expansion, is the Hubble parameter
$H(z)$.  It is easier to understand the physical meaning of the
dimensionless viscosity due to the Hubble expansion in terms of the
recession velocity of the gas due to the Hubble flow at the Bondi
radius: $v_{H}\sim r_B H(z)$. As expected, if $v_{H}/c_s=\hat
\beta_{Hubble} \ll 1$ we can neglect the cosmic expansion. The
dimensionless viscosity depends on the redshift, the mass of the PBH and
the temperature and ionization fraction of the cosmic gas:
\begin{eqnarray}
\hat \beta&=&\left({M_{pbh} \over 10^4 M_\odot}\right)\left({z+1 \over 1000}\right)^{3/2}\left({c_{s} \over 5.74~ {\rm km~s^{-1}}}\right)^{-3}\nonumber\\
&\times&\left[0.257+1.45\left({x_e \over 0.01}\right)\left({z+1 \over 1000}\right)^{5/2}\right].
\end{eqnarray}
\fig~\ref{fig:A} shows the accretion rate $\dot m$ from
equations (\ref{eq:acc1})-(\ref{eq:lambda1}) as a function of redshift. The
curves from bottom to top refer to $M_{pbh}=1, 10, 100, 10^3, 10^4,
10^5$ M$_\odot$. We assume a gas with constant
electron fraction $x_e=10^{-3}$ after recombination in the left panel and $x_e=10^{-1}$
in the right panel. Thermal feedback is neglected and recombination is
instantaneous.

If $f_{pbh}< 1$ we need to consider the effect of a growing dark
matter halo with mass $M_{h}(z)=\phi_i M_{pbh}[(1+z)/1000]^{-1}$. For
$\phi_i=3$ we find
\begin{equation}
{\dot m} = (0.016 \lambda) \left({1+z \over
    1000}\right) \left({M_{pbh} \over 1
    M_\odot}\right)\left({v_{eff} \over 5.74~{\rm km
    ~s}^{-1}}\right)^{-3}.
\label{eq:acc2}
\end{equation}
In the simulations presented in \S~\ref{sec:res}, we impose that the
dark halo stops growing when all the available dark matter has been
accreted; \ie, when $f_{pbh}\phi_i[(1+z)/1000]^{-1}= 1$.  The gas
accretion onto an extended dark halo of mass $M_h$ and power-law
density profile with logarithmic slope $2<\alpha<3$ is related to the
one for a point mass (with $M_{pbh}=M_h$) by the following scaling
relationships (see Paper~II):
\begin{equation}
\hat \beta^{(h)} \equiv \chi^{p \over 1-p}\hat \beta,~\lambda^{(h)} \equiv \Upsilon^{p \over 1-p} \lambda,~r_{cr}^{(h)} \approx \left({\chi \over 2}\right)^{p \over 1-p}r_{cr}.
\label{eq:lambda2}
\end{equation}
where $\chi=r_B/r_{h}< 2$, $p=3-\alpha$, and
\begin{equation}
\Upsilon = \left(1+10\hat{\beta}^{(h)}\right)^{1 \over 10}
\exp{(2-\chi)} \left({\chi \over 2}\right)^2.
\end{equation}
When the Bondi radius $r_B$ is larger than twice the radius of the
dark halo ($\chi \ge 2$), the dark halo behaves the same as a point
mass in terms of accretion rate, sonic radius and dimensionless
viscosity.  The numerical value of $\chi$ is
\begin{equation}
\chi=0.22 \left({M_{h}\over
  1 M_\odot}\right)^{{2 \over 3}}\left({1+z \over
  1000}\right)\left({c_{s} \over 1~{\rm km s^{-1}}}\right)^{-2}.
\end{equation}
We adopt the parameters $\alpha=2.25$, $\phi_i=3$ and $r_h$ given in
\eq~(\ref{eq:rhalo}), valid for a quasi-spherical dark halo in which
the dark matter does not fall directly onto the PBH.  Figure \ref{fig:B}
shows the accretion rate, ${\dot m}$, from
equations~(\ref{eq:acc2})-(\ref{eq:lambda2}) as a function of
redshift, with the curves representing the same masses as \fig~\ref{fig:A}
with the same simplifying assumptions going into the simulation.

\subsection{Accretion luminosity}\label{ssec:bol}

The radiative efficiency $\epsilon$ of accreting black holes is known
to depend on the accretion rate $\dot m$ and the geometry of the
accretion flow.  Observations and theoretical models suggest that the
radiative efficiency approaches a constant value $\epsilon \sim 0.1$
for accretion rates $\dot m \simgt 1$. If the accretion
rate is $\dot m \simlt 1$ the radiative efficiency is smaller by a
factor $\epsilon \propto \dot m$ \citep[\eg,][]{ParkO:01}. Depending
on the angular momentum of accreted material, a thick disk may
form. Otherwise the accretion geometry is spherical. The later case is
the most conservative as the accretion efficiency is minimum. In
addition, this case is appropriate for most PBH masses which have
$\dot m<1$ (see \S~\ref{ssec:ang}). We consider the following
accretion regimes:

\noindent
{\it Thin disk}: If $\dot m > 1$ we assume that the gas accreted by
the PBH forms a thin accretion disk and has radiative efficiency
$\epsilon = 0.1$. We assume that the bolometric luminosity cannot
exceed the Eddington limit and that, due to complex feedback effects
that we do not attempt to model here, the PBHs accrete near the
Eddington limit only for a fraction of time $f_{duty}$. Observations
of AGN at redshifts $z<3$ show that typically $f_{duty}$ is about
three percent \citep[\eg]{Shapley:03}. However, for PBHs, the duty
cycle can be substantially different as the nature of the feedback
mechanisms which regulate the cycle are not well understood. We let
$f_{duty}$ be a free parameter and we show that the limits on the PBH
abundance scale linearly with it. In summary,
\begin{equation}
l=f_{pbh} \times \min(0.1 {\dot m}, 1) ~~~{\rm if}~{\dot m}>1.
\label{eq:lacc_td}
\end{equation}
Figures \ref{fig:A}-\ref{fig:B} show that this choice of the radiative
efficiency is appropriate for PBHs with masses $M_{pbh}\simgt
100$~M$_\odot$.  The spectrum of the radiation emitted by the
accreting material depends on the accretion geometry, accretion rate,
and the mass of the black hole.  For a thin disk, any given radius
emits black body radiation with different emission temperature
(multicolor disk). The disk is hot in the inner parts and colder in
the outer parts. The maximum temperature of the disk depends on the
black hole mass: $T_{max}\propto M_{pbh}^{-1/4}{\dot m}$.  Beyond
photon energies $kT_{max}$ the emission is dominated by non-thermal
radiation produced by a diffuse hot corona around the accretion
disk. The spectrum of the non-thermal component is a power law of the
form $\nu L_\nu \propto \nu^{-\beta}$ with $\beta \sim 1.5$. Guided by
observations we model the spectrum as a double power-law with a break
at energies $h\nu = kT_{max}$ with slope $\beta=0.3$ at low energy and
$\beta=1.6$ at high energy.

\noindent
{\it Spherical accretion or ADAF:} If $\dot m < 1$ the accretion
efficiency depends on the flow geometry.  If the infall is
quasi-spherical the radiative efficiency is minimal.  The dominant
emission mechanism is thermal bremsstrahlung. Most of the radiation
will originate from the region just outside the event horizon. For the
case of accretion from a neutral \HI gas, including relativistic
effects, the efficiency for conversion of rest-mass energy into
radiation is $\epsilon=0.011 {\dot m}$ \citep{Shapiro:73,
Shapiro:73b}. Hence,
\begin{equation}
l=0.011 {\dot m}^2 ~~~{\rm if}~{\dot m}<1~{\rm (spherical~accretion)}.
\label{eq:lacc1}
\end{equation}
The spectrum is well approximated by a power law of the form $\nu L_\nu
\propto \nu^{0.5}$ at $\nu> 13.6$ eV and has an exponential cut-off at
$\nu \sim 5\times 10^5$ eV \citep{Shapiro:73}.  This case gives the
most conservative estimate of the effect of PBHs on the
cosmic ionization history. In addition some authors have found that
when the effects of magnetic field and cosmic rays are included in the
calculations, the radiative efficiency can be larger than the $0.01 \dot
m$ assumed in our fiducial case \citep{Meszaros:75b}.

If the gas has non-negligible angular momentum, an advection dominated
accretion flow (ADAF) may form \citep{Narayan:95}. In this case the radiative
efficiency is a factor of ten larger: $\epsilon = 0.1 \dot m$. Hence,
$l=0.1 {\dot m}^2$. This case is not particularly relevant for our
study as we have seen in \S~\ref{ssec:ang} that only PBHs with masses
$>1000$~M$_\odot$ can form accretion disks and typically for these
masses we find $\dot m>1$.

\subsubsection{Is the emerging radiation trapped by the accreting gas?}

Assuming spherical accretion and negligible pressure with respect to
the gravitational potential, the gas falling onto PBHs acquires a
velocity that approaches the free fall velocity near the black
hole.\footnote{For realistic cases in which the angular momentum of
accreting gas is important we refer to previous well known studies of
hot optically thin accretion flows
\citep{Shapiro:76,Narayan:95}.} From mass conservation it follows that
the gas density profile is $\rho_g \propto r^{-1.5}$ in the inner
parts. Hence, the gas column density $N_g \propto r^{-1/2}$ diverges
for radii $r \rightarrow 0$ and so does the Compton scattering optical
depth for outgoing radiation.

We checked whether the accreting gas may
become opaque to Compton scattering. In this case, the infalling gas would trap
the emerging X-ray photons (which are mostly emitted near the black hole
horizon) or the X-ray photons may be reprocessed into lower energy
photons.  The optical depth for emerging radiation emitted at a
distance $r_{min}=\alpha R_{Sch}$ from the black hole, 
\begin{equation} 
\tau_e ={\sigma_T \over m_p}\int_{r_{min}}^{r_{max}} \rho_g dr=
    {{\dot m}\over \alpha^{1/2}},
\end{equation}
where $\alpha \ge 1$ is the distance in units of the black hole
Schwarzschild radius and where we have assumed $r_{max} \gg
r_{min}$. Hence, for any value of the accretion rate smaller than
$\dot m \sim \alpha^{1/2} \sim 1$ the gas is optically thin to Compton
scattering. Since we assume that PBHs with $\dot m > 1$ form an thin
accretion disk, we can safely neglect trapping of radiation emitted
near the black hole in all the models we consider.  Similar
calculations show that the gas accreting onto PBHs is transparent (to
Compton scattering) to external background radiation.

\section{Feedback processes}\label{sec:feedback}

Gas accretion onto PBHs produces radiation that heats and ionizes the
IGM, thus raising its temperature above the value given in
\eq~(\ref{eq:temp}). X-ray photons, having a mean free path larger
than the mean distance between the sources, tend to build up a uniform
radiation background that increases the ionization fraction of the gas
and heats the IGM. The UV radiation, on the other hand, is absorbed
not very far from the emitting sources, so UV photons produce spheres
of fully ionized hydrogen around each PBH.  These processes introduce
global and local feedback effects which may alter the accretion rate.
Here we describe these effects and comment on their importance for 
the accretion calculation.

\subsection{Global Feedback}\label{ssec:global}

We use a semianalytic code, described in detail in \citep{RicottiO:03}
and based on \citep{Chiu:00}, to follow the chemical, ionization and
thermal history of the IGM from recombination to the redshift of the
formation of the first galaxies. We consider a gas of primordial
composition and include H$_2$ formation/destruction processes. Although
the code includes a stellar reionization and galaxy formation formalism,
in the present paper we focus on the cosmic epochs preceding the
formation of the first galaxies. In addition to the emission from PBHs
we include the contribution from the CMB radiation background that
gives us a redshift of recombination $z_{rec} \sim 1000$, in good
agreement with observations.

The thermal feedback by the X-ray
background is calculated self-consistently: the temperature and
ionization of the cosmic gas is determined by the luminosity of PBHs,
and the PBH accretion rate is a function of the gas temperature and
ionization. The thermal and chemical history is calculated iteratively
until the solution converges. The total emissivity per unit mass from
accretion onto PBHs is proportional to $l f_{pbh}$.  At $z<30$, when
the age of the universe is $t \sim 10^7$ yr, accretion onto PBHs stops
due to their peculiar velocities. Even if we assume that PBHs accrete
at the Eddington rate before $z=30$, the accretion time scale
$t_{Salp}=4.4 \times 10^8$ yr is much larger than the age of the
universe at this redshift. Hence, the growth of PBHs by gas accretion
is negligible.

\subsection{Local Feedback and Duty Cycle}

In addition to tracking the X-ray background, the semianalytic code
simulates the evolution of the Str\"omgren spheres produced by UV
radiation. These ionized bubbles have a small volume filling factor
and thus a negligible effect on the ionization history.  Nevertheless,
the gas is ionized and heated by UV radiation within the \HII regions,
leading to local feedback effects that may reduce the gas accretion
rate.

Local feedback may reduce the gas accretion rate if: i) the radius of
the \HII region exceeds the Bondi radius ($r_{HII} > r_B$), and ii)
if the gas temperature inside the \HII region is higher than the
temperature outside. Although this is always the case at low
redshifts, at $z>30$ the Compton coupling between the free electrons
inside the \HII regions and the CMB radiation is very effective in keeping
the gas temperature near the CMB temperature. 

If the final size of the \HII region is larger than the Bondi radius
($r_{HII}/r_B>1$) and the temperature inside the \HII region is much
larger than that outside, then the gas accretion rate may stop or
decrease. For simplicity, let's consider the most extreme scenario, in
which when $r_{HII}/r_B \ge 1$, the luminosity becomes negligible
and the \HII region recombines.  With this assumption the time
averaged luminosity can be estimated to be
\begin{equation}
\langle l \rangle_t={l \over 1+t_{off}/t_{on}}={l \over 1+(r_{HII}/r_B)^{1/3}}=f_{duty}l,
\label{eq:duty}
\end{equation}
where $t_{off}=t_{rec, H}$ is the \HI recombination timescale and
$t_{on}=t_{rec, H}(r_B/r_{HII})^{1/3}$ is the timescale it takes for the
ionization front to reach the Bondi radius $r_B$. Thus, it follows
$t_{off}/t_{on}=(r_{HII}/r_B)^{1/3}$. Equation~(\ref{eq:duty}) gives
a rough estimate of the duty cycle produced by local feedback.

\subsubsection{Temperature structure of the \HII region}\label{ssec:temp_hii}

We have performed time-dependent 1D radiative transfer simulations of
the \HII regions around PBHs to estimate $r_{HII}$ and the temperature
$T_{HII}$ inside the Str\"omgren radius. The basic results can be also
derived analytically. The radius of the Str\"omgren sphere around a
PBH of mass $M_{pbh}$ at the center of a spherical gas cloud with
density profile $n_{gas} = n_{H0}(r/r_0)^\gamma$, where $n_{H0}(z)$ is
the mean cosmic density of the gas, is
\begin{equation}
r_{HII} = (0.77~{\rm pc}) \left({S_{0,49}\over f_\gamma}\right)^{1/3} \left({1+z \over 1000}\right)^{-2},
\label{eq:stro}
\end{equation}
where $f_\gamma =1/3$ for $\gamma=0$ (\ie, assuming gas at uniform
density) or $f_\gamma \sim \ln(r_0/r_{min}) \sim 10$ assuming a 
free falling gas that becomes collisionally
ionized at $r_{min}$ (\ie, we adopt $\gamma=-1.5$ and $r_0/r_{min}
\sim 10^4$). We have defined the quantities
\begin{equation}
S_{0,49} \equiv (6 \times 10^{-3}) l \left({M_{pbh} \over 1~{\rm
M}_\odot}\right)\left({X \over 100}\right)^{-1},
\end{equation}
the number of ionizing photons emitted per second in units of $10^{49}$
photons s$^{-1}$, and
\begin{equation}
X \equiv \left({h \nu_0 \over L_{bol}}\int_{\nu_0}^\infty {L_\nu \over
    h \nu} d\nu \right)^{-1}
\end{equation}
the mean energy of the emitted photons in Rydberg units ($h \nu_0=13.6$
eV). Assuming a power-law spectrum with index $\beta=0.5$ in the
frequency interval $13.6 < \nu < 5 \times 10^5$ eV, appropriate for
spherical accretion, we find $X \sim 192$. For thin disk accretion we
find $X=\beta/(\beta-1) \sim 3$ for $\beta=1.5$. In order to estimate
the value of the Str\"omgren radius we assumed an \HI recombination
coefficient $\alpha=1.28 \times 10^{-12}$ cm$^{-3}$s$^{-1}$, appropriate
for a gas at $T \sim 2,000$ K. This temperature is typical for the gas
inside an \HII region at $z \sim 500$.

We now estimate the temperature inside the \HII region assuming
that the dominant gas heating is \HI photoionization and the dominant
cooling is Compton cooling. In a uniform density gas in ionization and
thermal equilibrium the temperature inside the \HII region is
constant. At equilibrium the ionization rate per hydrogen atom equals
the recombination rate: $t_{ion=}t_{rec,H} \sim (124~{\rm
yr})[(1+z)/1000]^{-3}$ and the heating rate is $\langle h \nu
\rangle/t_{ion}$, where $\langle h \nu \rangle \sim 0.365~{\rm
eV}/(\beta+2)$ is the mean photon energy deposited into the gas per
hydrogen atom. Neglecting Compton heating, which is only important in a
small volume around the black hole, a fraction $C \sim
0.0268/(\beta+2)X$ of the radiation emitted by the black hole is
deposited as heat inside the \HII region. If we assume that the
heating rate equals the Compton cooling rate
$k(T_{HII}-T_{CMB})/t_{Comp}$, where $t_{Comp} \sim (0.76~{\rm
yr})[(1+z)/1000]^{-4}$ we find
\begin{equation}
{T_{HII} \over T_{CMB}} \approx 1+{0.36 \over \beta+2}\left({1+z \over 1000}\right)^{-2}.
\label{eq:thii}
\end{equation}
Thus, for spherical accretion with $\beta=0.5$ (disk accretion with
$\beta=1.5$) the temperature inside the \HII regions at $z \sim 500$
is only 56\% (41\%) higher than the temperature outside, which is
roughly equal to the CMB temperature $T_{CMB} \approx 1375$. We
conclude that at high redshift, even if the Bondi radius is smaller
than the Str\"omgren radius, the increase of the IGM temperature due
to local feedback is negligible. Hence, we can neglect the local
contribution to thermal feedback (from UV photos) with respect to the
global thermal feedback from X-ray heating, that is always included
self-consistently (see \S~\ref{ssec:global})
\begin{figure}[t]
\epsscale{1.1}
\plotone{\figname{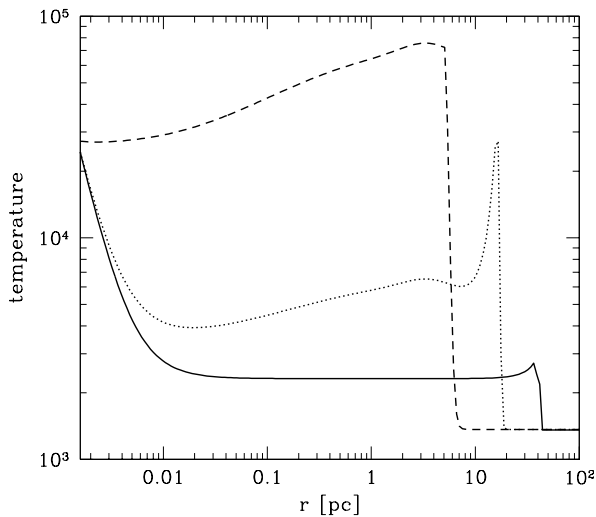}}
\caption{(a) Temperature structure of the \HII region around a PBH at
  $z=500$ emitting $S_0=10^{52}$ ionizing photons per second. The
  curves show the temperature profile after $2$ yr (dashed), $100$
  yr (dotted) and $4600$ yr (solid) after the source turns on. The
  source spectrum is one appropriate for spherical accretion onto a
  black hole, with log-slope $\beta=0.5$.}
\label{fig:temp_str}
\end{figure}

More sophisticated radiative transfer simulations confirm the results
of the analytical calculations of $r_{HII}$ and $T_{HII}$. In
\fig~\ref{fig:temp_str} we show the time evolution of the temperature
profile inside an \HII region around a PBH with $S_{0}=10^{52}$ s$^{-1}$ at
$z \sim 500$. The 1D radiative transfer code used to produce the
temperature plots is described in detail in \cite{RicottiGS:01}. In
addition, in the present simulation we have also included Compton
heating from X-rays emitted near the PBH. The increase of the gas
temperature at small radii evident in \fig~\ref{fig:temp_str}, $T
\propto r^{-2}$, is produced by Compton heating. We estimate that
Compton heating becomes dominant over photoionization heating at radii
\begin{eqnarray}
r &<& \left({X \over \langle h\nu\rangle}\right)^{1 \over 2} \left({kT \over
  m_ec^2}\right)^{1 \over 2} (\sigma_T S_0
t_{rec,H})^{1 \over 2} \nonumber\\
&\approx& (0.01 ~{\rm pc}) S_{0,49}^{1 \over 2}
  \left({1+z \over 1000}\right)^{-{3 \over 2}}.
\end{eqnarray}
After a time-dependent phase during which the \HII region reaches its
Str\"omgren radius, the temperature inside the \HII region decreases
to a constant value $T_{HII} \sim 2,100$ K, about $56\%$ higher than
the temperature outside the \HII regions, in agreement with the
analytical estimate. We have verified that the temperature $T_{HII}$ is
independent of the source luminosity and depends on $\beta$ and the
redshift according to \eq~(\ref{eq:thii}).

\subsubsection{Str\"omgren vs Bondi radii}

Let's now estimate the ratio $r_{HII}/r_B$, which, along with the HII
region gas temperature, determines whether local feedback will reduce
the gas accretion rate.
>From \eq~(\ref{eq:bondi})
and \eq~(\ref{eq:stro}) with $f_\gamma=10$ we find
\begin{eqnarray}
{r_{HII} \over r_B} &\sim& 3\times 10^3 \left({l \over X}\right)^{1 \over
  3}\left({M_{pbh} \over 1~M_\odot}\right)^{-{2 \over 3}}\nonumber\\
&\times& \left({1+z \over 1000}\right)^{-2} \left({v_{eff} \over {\rm 5.7 km~s^{-1}}}\right)^2
\end{eqnarray}
for ``naked'' PBHs. If we include the growth of an extended dark halo,
${r_{HII}/r_B}$ decreases by a factor $0.3[(1+z)/1000]$ at
$z<1000$. 
We shall consider two cases for the accretion luminosity: Case~A,
appropriate for small accretion rates $\dot m<1$ and Case~B,
appropriate for $\dot m \sim 1-10$ (see \S~\ref{ssec:bol}).

{\it Case A:} $l \propto {\dot m}^2$.

\noindent
The case $l \propto \dot m^2$ is the most interesting because is the
most likely to occur for PBHs and because the ratio $r_{HII}/r_B$ is
independent of the gas effective velocity $v_{eff}$ and PBH
mass. Indeed, $r_{HII} \propto l^{2/3} \propto v_{eff}^{-2}$ and
$r_B \propto v_{eff}^{-2}$. Assuming $\epsilon=0.01$ (spherical
accretion) we find that the ratio
\begin{equation}
{r_{HII} \over r_B} \sim 2 \left({X \over 200}\right)^{-{1 \over 3}}
\end{equation}
is a constant of order of unity for a ``naked'' PBH with mass
$M_{pbh}<M_{cr} \sim 10^4$ M$_\odot$ and ${\dot m}<1$. For PBHs more
massive than $M_{cr}$, the constant is smaller by a factor
$(M_{pbh}/M_{cr})^{2/3}$, thus local radiative feedback can be safely
neglected because $r_{HII} < r_B$. Including the growth of the dark
halo around PBHs we find:
\begin{equation}
{r_{HII} \over r_B}= 0.66\left({1+z
  \over 1000}\right)\left({X \over 200}\right)^{-{1 \over
  3}} < 1,
\label{eq:rhii}
\end{equation}
valid only at redshifts $z \simgt 100$. At lower redshifts, $r_{HII}/r_B$
increases with respect to the estimate in \eq~(\ref{eq:rhii}) but
its value remains on order unity or smaller for $\dot m <1$.

{\it Case B:} $l \propto \dot m$.

\noindent
This case applies to the case of relatively massive PBHs with typical masses
$>10^3$ M$_\odot$ that have $1<\dot
m<10$ or $0.1<l<1$. It is sufficient to consider the
case of a PBH clothed in its dark halo. We find
\begin{equation}
{r_{HII} \over r_B} \sim 4 \left({M_{pbh} \over
10^4 M_\odot}\right)^{-2/3} \left({X \over
  3}\right)^{-1/3}.
\end{equation}
Thus, $r_{HII}< r_B$ for $M_{pbh} \simgt 10^5$ M$_\odot$. Also in this
case we can neglect local feedback with the exception of the mass
range $10^3-10^5$ M$_\odot$ where local thermal feedback may reduce
accretion by a factor of a few (from \eq~(\ref{eq:duty}) we obtain a
reduction $f_{duty} \sim 30\%$ for $M_{pbh} \sim 10^3$
M$_\odot$).

In summary, the local feedback due to the formation of an \HII region
around the PBH can be neglected in most cases. At high redshift the
temperature inside the \HII regions is only a few tens of a percent
higher than the temperature outside (see \eq~[\ref{eq:thii}]), thus
even if the Bondi radius is smaller than the Str\"omgren radius, the
reduction of the accretion rate due to thermal feedback is negligible.
At lower redshifts ($z<100$) the temperature inside \HII regions is
larger than the temperature outside but we found that the Str\"omgren
radius is typically smaller than the Bondi radius.

The only exceptions are massive PBHs with $10^3 M_\odot< M_{pbh} <10^5$
M$_\odot$ for which we estimate $f_{duty} \simlt 30\%$.  For this
case, since $r_{HII} > r_B$, we need to also assume that the gas is
fully ionized at the Bondi radius. Hence, in addition to thermal
feedback, the ionized gas inside the \HII region increases the gas
viscosity due to Compton drag. Compton drag is negligible at $z<100$
(see Figs.~\ref{fig:A}-\ref{fig:B}) but reduces the accretion rate at
$z>100$ when, instead, thermal feedback is negligible.

\section{Simulating the cosmic ionization}\label{sec:res}
\begin{figure*}[t]
\epsscale{1.15}
\plottwo{\figname{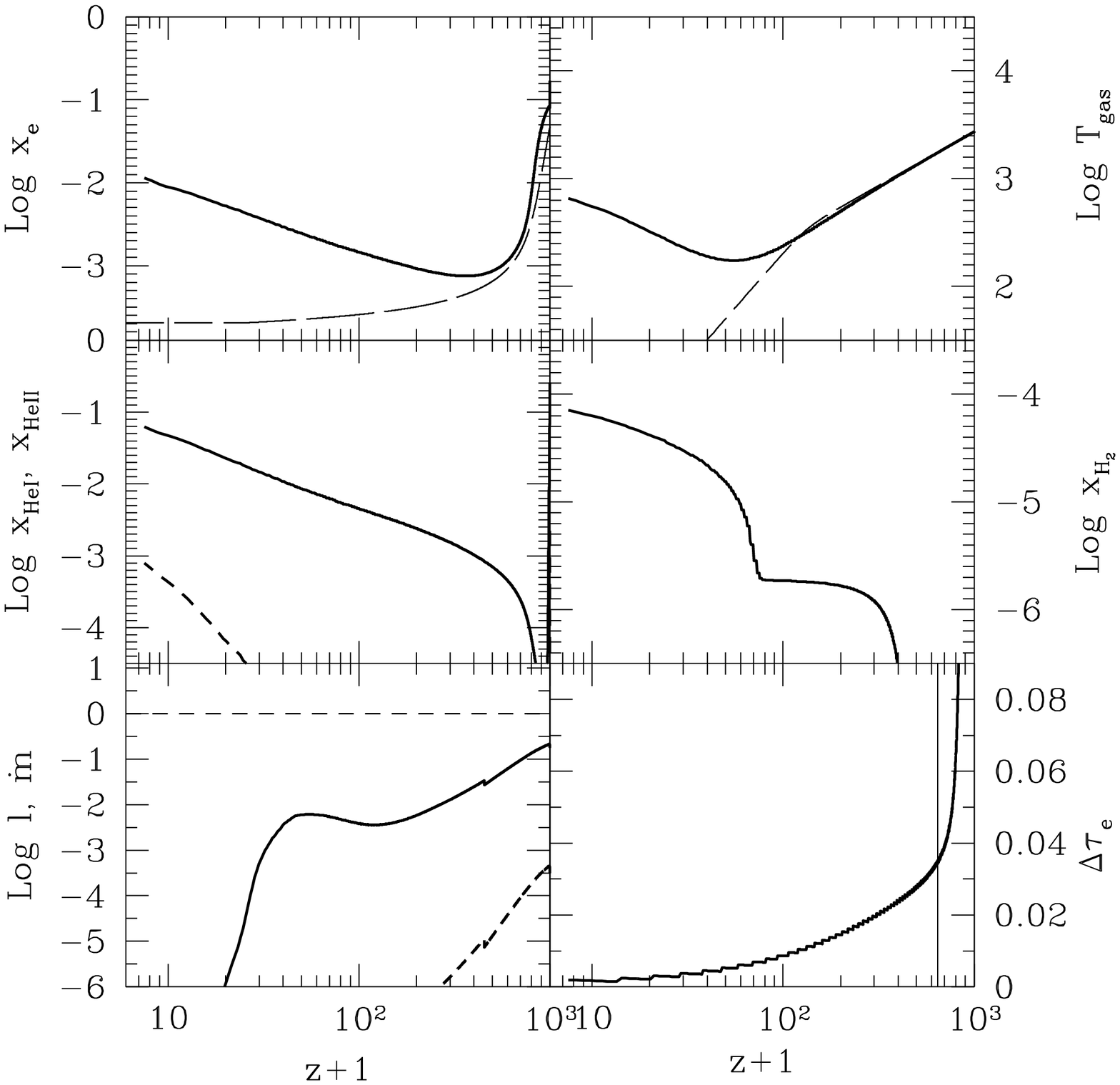}}{\figname{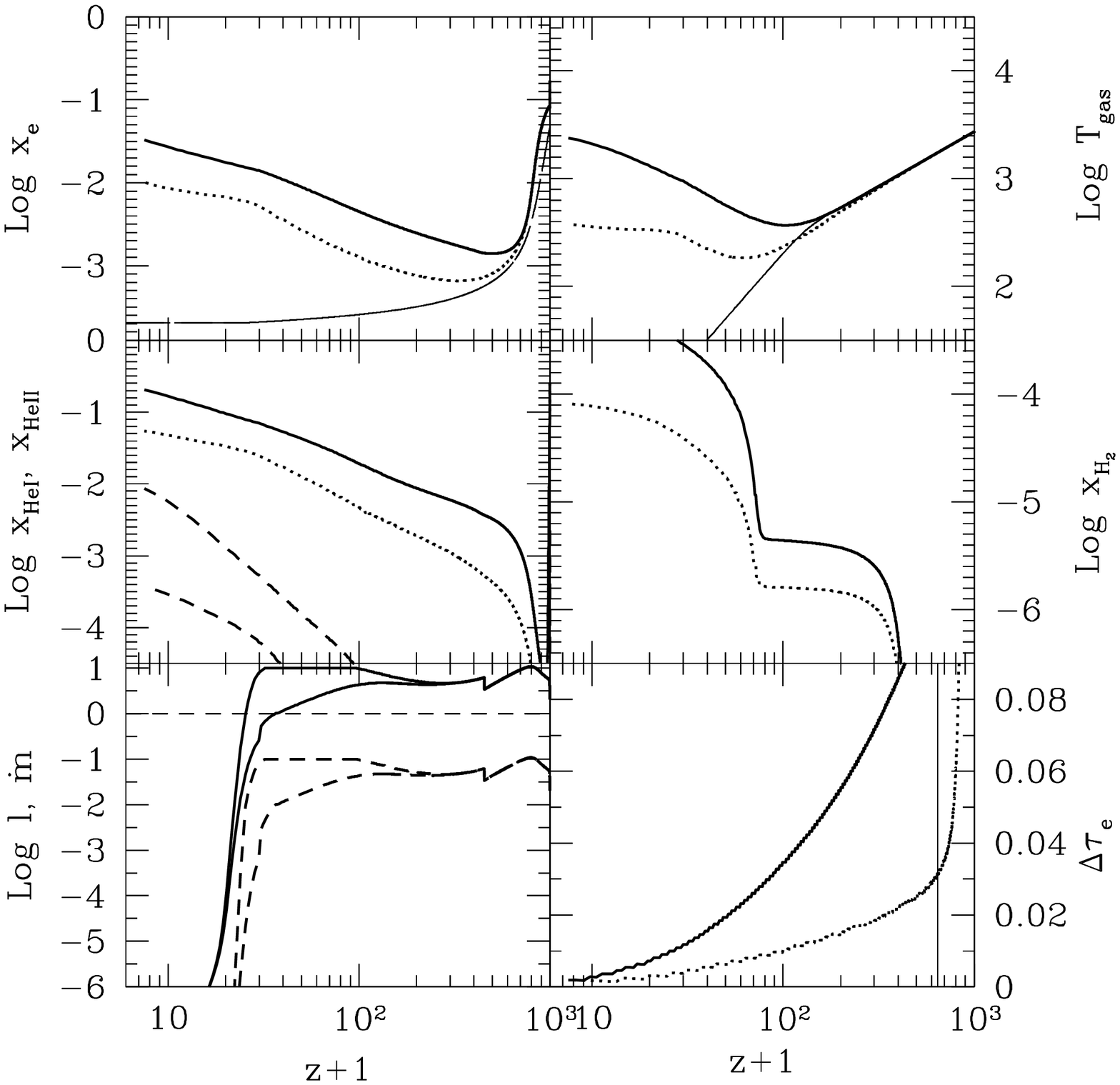}}
\caption{{\it Left}. Simulation of early ionization by PBHs with mass
  $M_{pbh}=100$ M$_\odot$ and abundance $f_{pbh}=10^{-4}$. The panels
  show: (a) the ionization fraction $x_e$ (the dashed line shows the
  recombination history without PBHs); (b) the temperature (the dashed
  line shows the thermal history without PBHs); (c) the He ionization
  fraction; (d) the H$_2$ abundance; (e) the accretion rate, $\dot m$
  (solid curve) and luminosity $l$ (dashed curve); (f) the optical
  depth to Thompson scattering $\tau_e$. {\it Right}. Same as in the left
  panel but for $M_{pbh}=1000$ M$_\odot$ and $f_{pbh}=10^{-6}$ (solid
  curves) and $f_{pbh}=10^{-7}$ (dotted curves).}
\label{fig:rei1}
\end{figure*}

We simulate the ionization, chemical and thermal history of the
universe after recombination using a modification of the semianalytic
code discussed in detail in \cite{RicottiO:03}. The original code is
tailored to simulate UV and X-ray reionization from high redshift
galaxies at $z < 30$. Here we do not include any ionizing source other
than PBHs. In order to simulate standard recombination properly we
model the redshifted CMB black body radiation including the effect of
Thompson opacity. We test the recombination history calculation in the
absence of PBHs against a widely used code RECFAST \citep{Seager:99}.
We can reproduce the recombination history at redshifts $z < 900$
(when $x_e<0.1$) with accuracy of a few percent. Using the equations
derived in this work we model the UV and X-ray emission from accreting
PBHs including feedback effects, the secondary ionization due to fast
photo-electrons, and Compton heating/cooling, and we solve the
chemical network for a gas of primordial composition (\eg, molecular
hydrogen formation/dissociation). We run a grid of models with a range
of PBH masses spanning several orders of magnitude, between $10^{-3}$
M$_\odot$ to $10^8$ M$_\odot$.

For the sake of simplicity, throughout the rest of the paper we
consider the effects of a population of equal mass PBHs which account
for a fraction $f_{pbh}$ of the dark matter. The results can be used
to estimate the effect of an arbitrary PBH mass function by
integrating the optical depth $d \tau_e(M)$ in each mass bin with
weight function $(df_{pbh}/dM)^{1/2}$. See below for an explanation
of the choice of the weighting function. Special care should be taken
in excluding from the integration PBHs which are sufficiently small to
be part of the dark halos of more massive PBHs. In addition we need to
check that that the proper velocities of small PBHs, due to their
interaction with more massive ones, is negligible (\ie,
subsonic). Fitting formulas for $\tau_e$ as a function of $M_{pbh}$
and $f_{pbh}$ are provided at the end of this section.

Independently of the radiation emitted by the gas falling directly
onto the PBHs, gas inflow onto the PBH can produce shocks and
collisional ionization. For the idealized case of an isothermal gas
the sonic point is located at $0.5r_B$, where $r_B$ is the Bondi
radius. At this point a shock develops that ionizes the gas, even
if the temperature remains below 10,000 K due to Compton cooling and
molecular hydrogen cooling. The volume filling factor of the
collisionally ionized gas is $f_{V} \sim n_{pbh}(r_B/2)^3$, where
$n_{pbh}$ is the PBH physical number density. Assuming that a fraction
of the IGM $f_{V}$ is fully ionized and the rest of the cosmic gas is
neutral we have a volume-weighted ionization fraction due to shocks:
$\langle x_e \rangle_V \approx 10^{-11} f_{pbh} ({M_{pbh}/1~{\rm
M}_\odot})^2$.  This effect is negligible for any PBH mass assuming
values of $f_{pbh}$ allowed by observational constraints.

Examples of simulations with $M_{pbh}=100$~M$_\odot$ and
$1000$~M$_\odot$ are shown in \fig~\ref{fig:rei1}.  Here, we describe
simulations in models that are consistent with FIRAS and WMAP3
data. In these models the contribution of PBHs to the total optical
depth to Thompson scattering typically does not exceed $\Delta \tau_e
\sim 0.03-0.04$ at 68\% confidence level and $\Delta \tau_e \sim
0.07-0.09$ at 95\%confidence level.  The contribution is uncorrelated
with the value of $\tau_e$ produced by ionization sources in high
redshift galaxies (at $z<30$) which is $\tau_e=0.09 \pm 0.03$ (see
\S~\ref{sec:obs}).

The mean electron fraction increases approximately as $x_e \propto
(z+1)^{-1}$ from $x_e \sim 10^{-3}$ at $z \sim 900$ to values close to
$x_e \sim 10^{-1}-10^{-2}$ at $z \sim 10$. Hence, the viscous effect
of Compton drag can be neglected after recombination. The main
contribution to the partial ionization of the cosmic gas after
recombination is from X-ray emission at redshifts $z > 100$. At
smaller redshifts, due to the rapid decline of the accretion rate onto
PBHs, the ionization fraction keeps increasing slowly, due to
redshifted X-ray background photons \citep{RicottiO:03}.  The gas
temperature is approximately equal to the CMB temperature at $z>200$
and ranges between $100$ K to $1000$ K afterwards. The \GII ionization
fraction becomes $1-10$\% and the molecular abundance $x_{H_2} \sim
10^{-4}-10^{-5}$ after redshift $z \sim 100$. The molecular abundance
is 10-100 times larger than the standard value, $x_{H_2} \sim
10^{-6}$, obtained neglecting PBHs.  Depending on the mass of PBHs we
encounter different regimes for the accretion.

\noindent
1) $1~{\rm M}_\odot < M_{pbh}< 30$~M$_\odot$ : The dark halo has
   negligible effect on the accretion rate. The accretion is spherical
   and the luminosity $l \propto {\dot m}^2$. Global feedback effects
   are negligible.

\noindent
2) $30~{\rm M}_\odot < M_{pbh}< 300$~M$_\odot$ : The dark halo
   increases the accretion rate onto PBHs, mostly at redshifts
   $z<100$. The accretion is spherical
   and the luminosity $l \propto {\dot m}^2$. Global feedback effects
   are important in reducing the accretion rate at $z<100$.

\noindent
3) $300~{\rm M}_\odot < M_{pbh}< 3000$~M$_\odot$: The dark halo
   increases the accretion rate onto PBHs at all redshifts. The
   accretion rate is large (${\dot m} \simgt 1$), thus a thin disk may
   form and the luminosity is $l \propto f_{duty}{\dot m}$. Global
   feedback effects moderately reduce the accretion rate.

\noindent
4) $M_{pbh}> 3000$~M$_\odot$ : The dark halo increases the accretion
   rate onto PBHs at all redshifts. For masses $>5\times
   10^4$~M$_\odot$, the Hubble expansion becomes important and the
   Bondi solutions transition to self-similar infall solutions (see
   Paper~II). PBHs accrete at the Eddington rate with a duty cycle
   $f_{duty}$. Global feedback effects do not reduce the accretion
   rate significantly for values $\Delta \tau_e \simlt 0.05$,
   compatible with WMAP3 (see \S~\ref{sec:obs}).

The qualitative dependence of $x_e(z)$ and $\tau_e(z)$, on the mass
and abundance of PBHs can be understood assuming ionization
equilibrium. For $x_e(z) \ll 1$ and neglecting the temperature
dependence of the recombination coefficient we have $x_e(z)^2 \propto
l(z, M_{pbh}) f_{pbh}$. For PBHs with masses smaller than $M_{pbh}
\sim 100$ M$_\odot$ we have $\dot m <1$ and $l \propto
M_{pbh}^2$. Hence, $\tau_e \propto x_e \propto M_{pbh} f_{pbh}^{1/2}$.
If the mass of the PBH is $ \simgt 100$ M$_\odot$ the accretion rate
is typically $\dot m >1$, $l \propto M_{pbh}$ and $\tau_e \propto x_e
\propto (M_{pbh} f_{duty}f_{pbh})^{1/2}$. Finally, if $M_{pbh} \simgt
1000$ M$_\odot$ we have $\dot m \ge 10$, PBHs accrete nearly at the
Eddington limit and $\tau_e \propto x_e \propto
(f_{duty}f_{pbh})^{1/2}$ is independent of the PBH mass.  To first
order, the value of $\Delta \tau_e$ produced by PBHs of mass $M_{pbh}$
and abundance $f_{pbh}$ can be parameterized as follows:
\begin{equation}
\Delta \tau_e \approx
\cases{
0.05 \left({M_{pbh}\over 1~M_\odot}\right)f_{pbh}^{1 \over 2}&
  {if~$M_{pbh} <100$ M$_\odot$,}\cr
0.1 \left({M_{pbh}\over 1~M_\odot}\right)(f_{duty} f_{pbh})^{1 \over 2}&
  {if~$10^2 < {M_{pbh} \over 1~M_\odot}<10^3$,}\cr
10^{5} (f_{duty} f_{pbh})^{1 \over 2}& {if~$M_{pbh}>10^3$ M$_\odot$.}\cr
}
\label{eq:deltataue}
\end{equation}
The fit is derived from a grid of simulations with $\Delta \tau_e <
0.2$ and is not accurate for large values of $\Delta \tau_e$ due to
feedback and saturation effects.

\section{Effects of PBHs on the CMB spectrum and anisotropies}\label{sec:obs}

\subsection{Temperature and Polarization Anisotropies}

\begin{figure}[t]
\epsscale{1.2}
\plotone{\figname{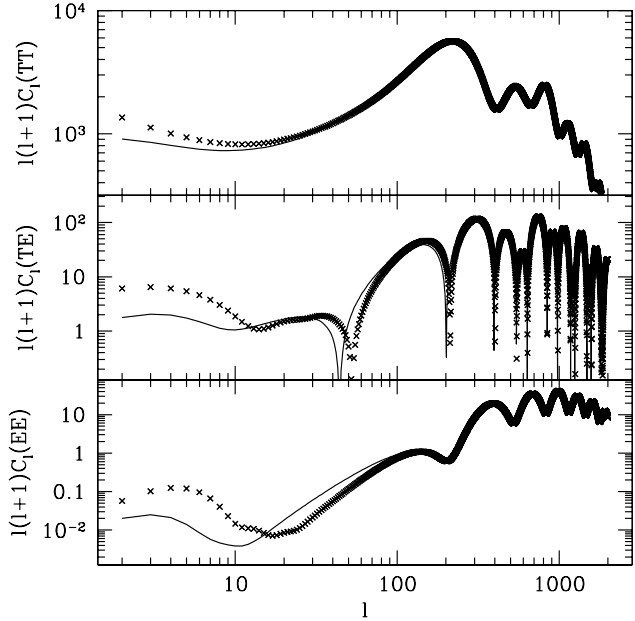}}
\caption{Power spectrum of temperature and polarization anisotropies
  for the best fit WAMP3 model with $\tau_e =0.09$ and redshift of
  reionization $z_{rei}=11$ (crosses) and a model with the same
  $\tau_e$ but $z_{rei}=7$ and modified recombination history with
  constant residual ionization fraction $x_{e0}=2.5 \times
  10^{-3}$. The panels show from top to bottom: TT, EE and TE power
  spectra.}
\label{fig:TT}
\end{figure}

In \S~\ref{sec:res} we have shown that a signature of the existence of
PBHs is the modification of the cosmic recombination history. It is
possible to construct models with $\tau_e \sim 1$ due to the large
residual fractional ionization of the cosmic gas after recombination
produced by X-ray ionization. Of course, such models are ruled out by
CMB observations. The WMAP3 constraint on $\tau_e$ is $\tau_e \sim
0.09\pm 0.03$. This limit on $\tau_e$ assumes that the intergalactic
medium becomes partially or fully ionized by stars and/or black holes
starting at redshift $z \simlt 30$. As illustrated in
\fig~\ref{fig:TT}, the partial ionization from PBHs and the ionization
from high redshift galaxies forming at $z<30$ produce very different
signatures on the CMB polarization anisotropies. PBHs affect small
angular scales with $l \simgt 10$, while radiation emitted by high
redshift galaxies affect larger scales with $l<10$.  Hence, PBHs do
not contribute significantly to the value $\tau_e=0.09$ quoted by the
WMAP3 team, which must be produced by ionization sources other than
PBHs.

In order to constrain models which allow for the existence of PBHs we
include an additional cosmological parameter describing the deviation
from the standard recombination history calculated using RECFAST
\citep{Seager:99}. We have modified the publicly available codes
CAMCMB and COSMOMC \citep{Lewis:02} to include an additional degree of
freedom.  The redshift dependence of the ionization history after
recombination depends weakly on the mass of PBHs for values of $\Delta
\tau_e \simlt 0.1$. We modify the recombination history $x_{e,
rec}(z)$ given by RECFAST as follows:
\begin{equation}
x_e(z)=x_{e, rec}(z) + \min\left[x_{e0} \left({1+z \over 1000}\right)^{-1}, 0.1\right],
\label{eq:modelA}
\end{equation}
The free parameter $x_{e0}$ is the ionization fraction at
$z=1000$. The ionization fraction increases proportionally to the
scale parameter after recombination and is constant at later times.
In addition, we model the instantaneous reionization at $z=z_{rei}$
produced by galactic sources. We will show that the effect on the CMB
polarization anisotropies of PBHs and other ionizing sources are
uncorrelated.

\begin{figure*}[t]
\epsscale{1.1}
\plottwo{\figname{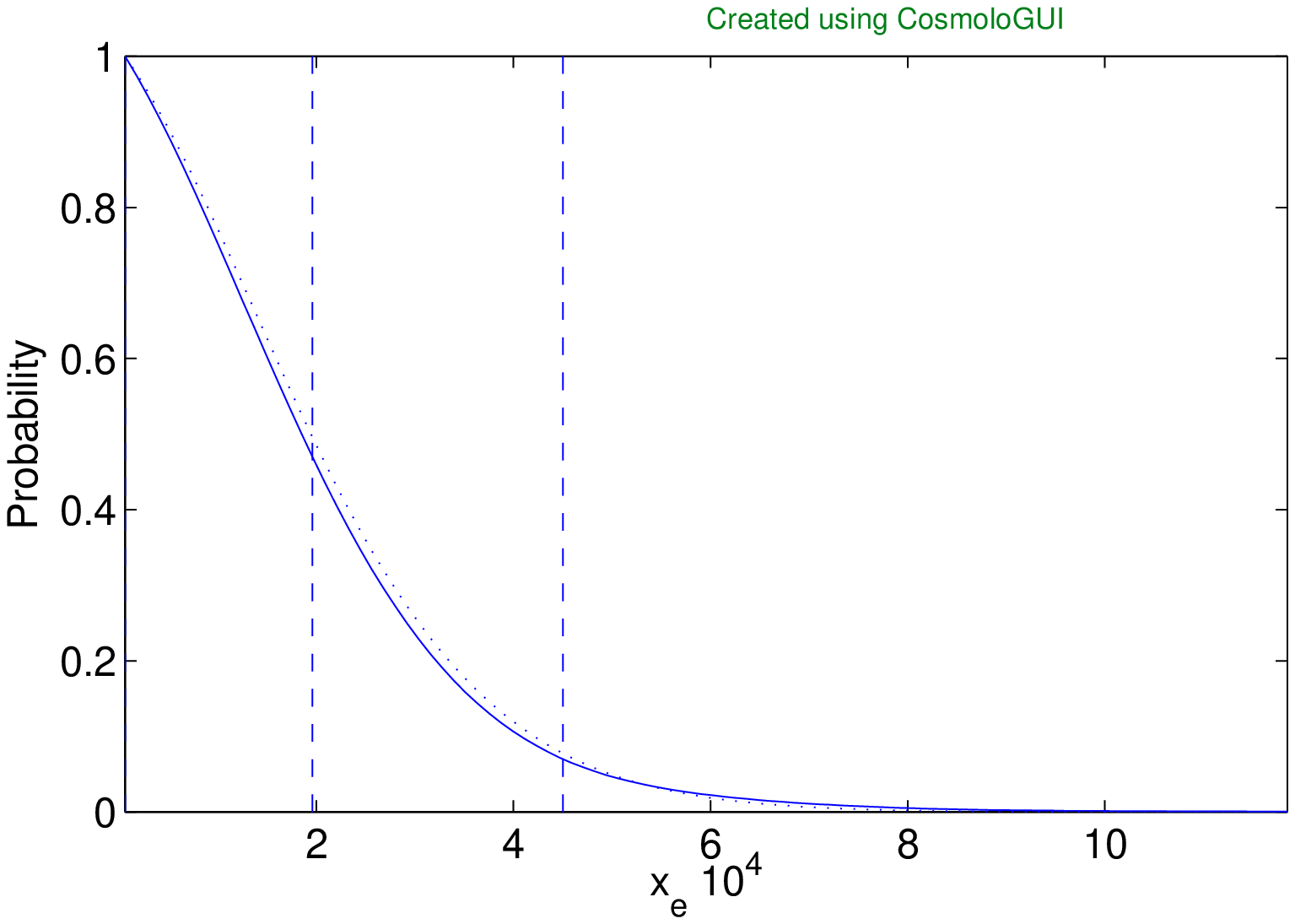}}{\figname{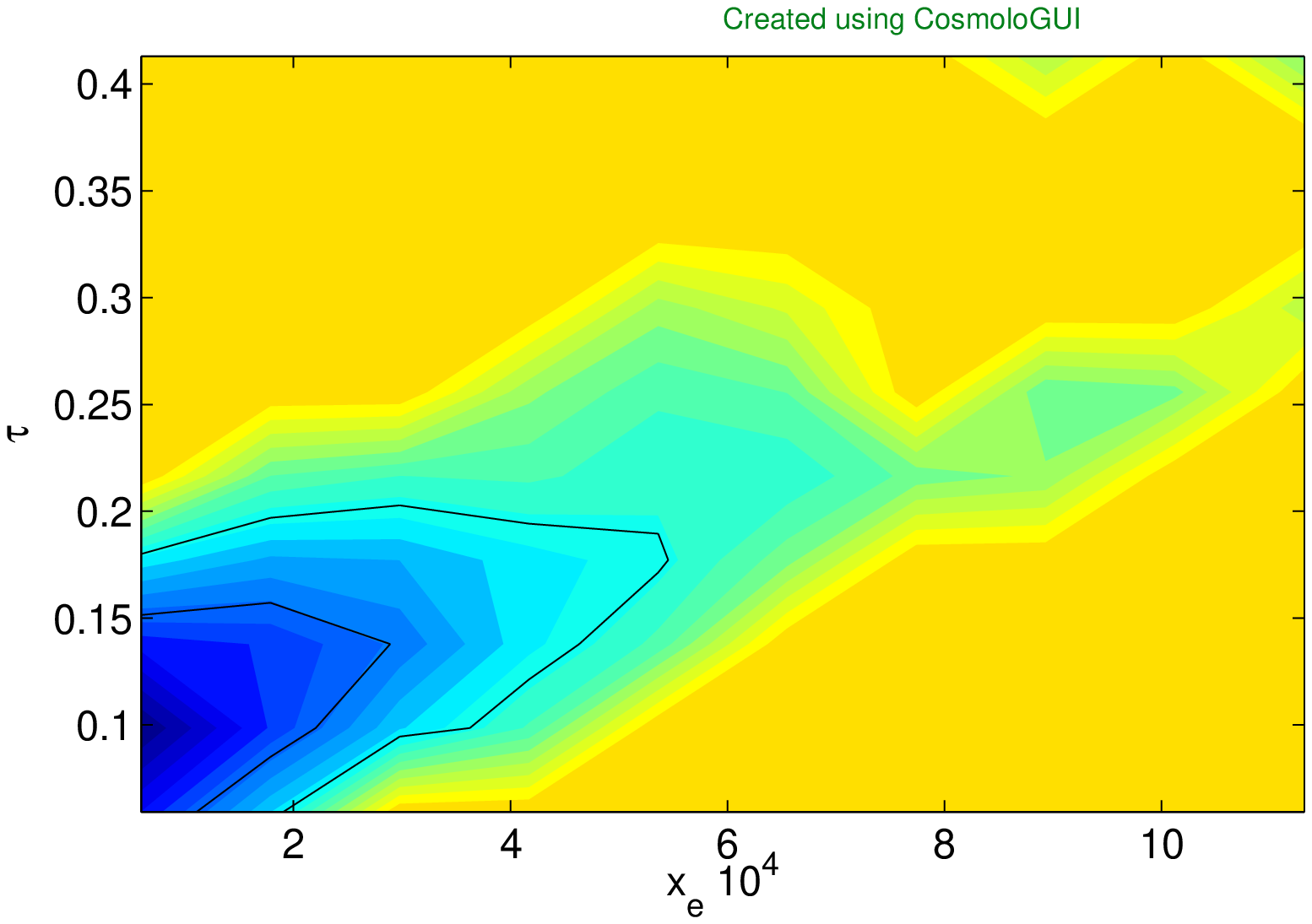}}
\epsscale{1.1}
\plottwo{\figname{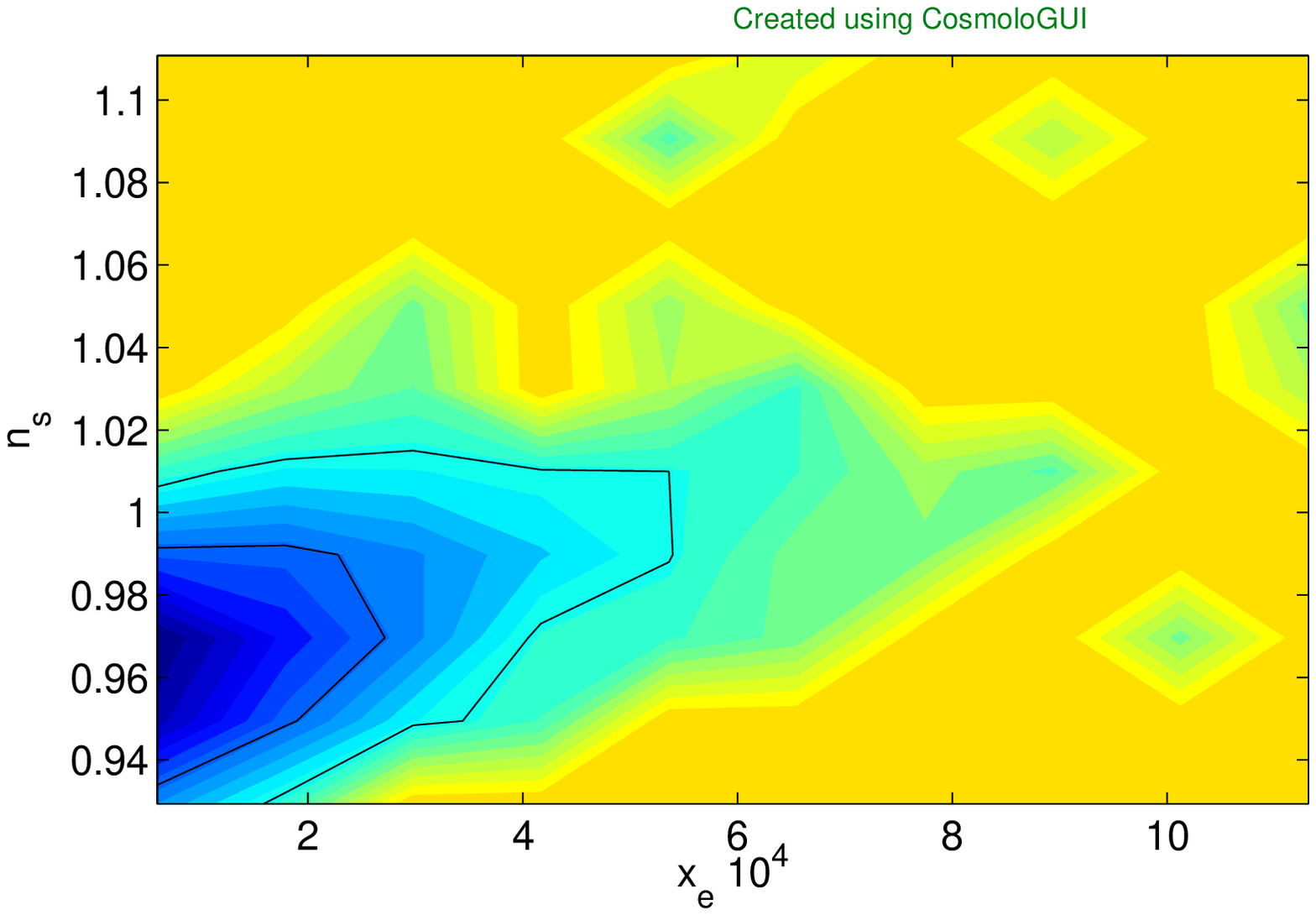}}{\figname{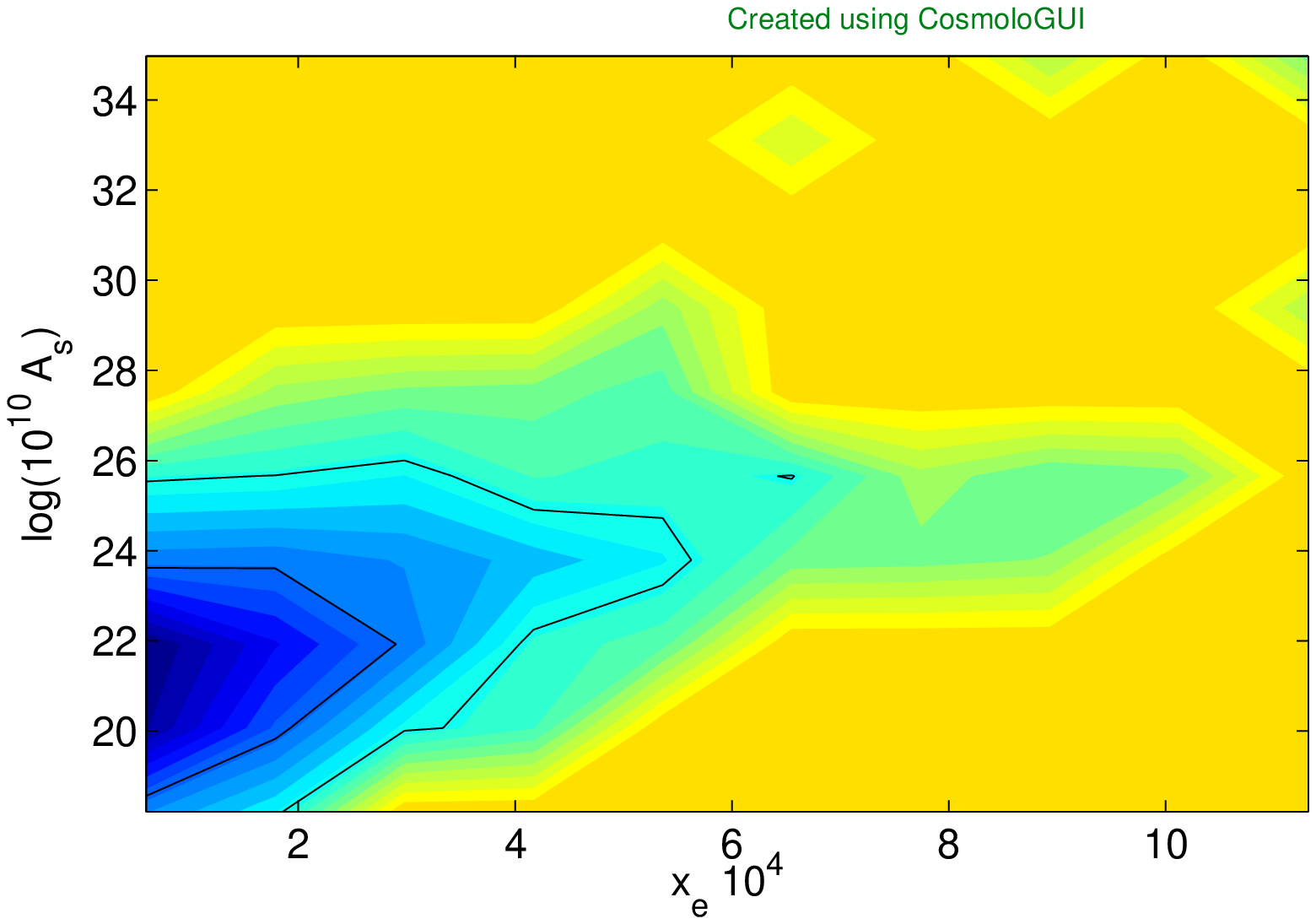}}
\caption{Likelihood isocontours of cosmological parameters as a
  function of $x_{e0}$ which describes the modified recombination
  history produced by PBHs (see the text). We have used WMAP3 data and
  a modified version of COSMOMC \protect{\citep{Lewis:02}}. The equation used
  for the non-standard recombination history is given in
  \protect{\eq~(\ref{eq:modelA})} and the free parameter $x_{e0}$
  describes the electron fraction at $z=1000$. The four panels from
  the top-left corner show (clockwise): i) marginalized likelihood of
  $x_{e0}$ (in units of $10^{-4}$) with 68\% and 95\% confidence
  limits; ii) optical depth $\tau_e$ vs $x_{e0}$; iii) amplitude of
  the power spectrum vs $x_{e0}$; iv) scalar tilt of the initial power
  spectrum depth $n_s$ vs $x_{e0}$.}
\label{fig:cmb_par1}
\end{figure*}

Using the Markhov Chain Monte-Carlo code COSMOMC with WMAP3 data we find
$1\sigma$ and $2\sigma$ marginalized confidence limits for the new
free parameter: $x_{e0}< 2 \times 10^{-4}$ (68\% CF)  and $x_{e0}< 4.2
\times 10^{-4}$ (95\% CF). These upper limits on the fractional
ionization correspond to upper limits on $\Delta \tau_e=\tau_e -
\tau_{e, rei}$ of $0.05$ (68\% CF) and $0.1$ (95\% CF).
As illustrated in \fig~\ref{fig:cmb_par1}, the new parameter does not
correlate with the redshift of reionization $z_{rei}$ but because it
increases the total $\tau_e$ it correlates with $\sigma_8$ and $n_s$,
increasing their values with respect to the ones quoted by the WMAP3
team \citep{Spergel:06}. Invoking a non-standard recombination history
may ease the tension between the low value of $\sigma_8 \sim 0.74$
from the WMAP3 analysis (assuming standard recombination history) and
clusters data that instead seem to favor larger values of $\sigma_8
\sim 0.9$ \citep{Evrard:07}, but see \citep{BodeO:07}.

Using the results of the simulations presented in \S~\ref{sec:res}
that are approximately summarized by \eq~(\ref{eq:deltataue}), we are
able to constrain the abundance of PBHs with masses $1~{\rm
M}_\odot < M_{pbh}<10^8$~M$_\odot$. The results are shown by the thick
solid line in \fig~\ref{fig:FIRAS}(left).

\subsection{Spectral Distortions}

The injection of some form of energy density, $\Delta U$, in the
expanding universe at redshifts $z<10^7$ cannot be fully thermalized,
and so it produces observable deviations of the CMB spectrum from a
perfect black-body \citep[\eg,][]{Burigana:91}. There are two types of
spectral distortions: $\mu$ and $y$-distortions. Energy injection in
the redshift range $10^7<z<10^5$ produces a Bose-Einstein spectrum
with chemical potential $\Delta U/U =0.71 \mu$, where $U$ is the
unperturbed energy density of the CMB. In the redshift interval $10^5
< z < 1000$ the Comptonization distortion dominates and has amplitude
$\Delta U/U = 4y$. Using the spectrometer FIRAS on the COBE
satellite, \cite{Fixsen:96} found the following upper limits for the
deviation of the CMB from a Planck spectrum: $\mu \le 9 \times 10^{-5}$
and $y \le 1.5 \times 10^{-5}$ at at 95\% confidence.  It follows that
$\Delta U/U \le 6 \times 10^{-5}$ at 95\% confidence from
$10^7<z<1000$.  In this section we calculate the value of the
$y$-parameter produced by a population of PBHs of mass $M_{pbh}$ and
compare it to observed upper limits from FIRAS to set upper limits on
their abundance.  In order to calculate the total value of the
$y$-parameter we need to distinguish between three epochs for the
energy injection:

\noindent
1) Before the redshift of last scattering at $z_{rec} \sim 1000$ the
universe is optically thick to Compton scattering hence all the energy
emitted by accreting PBHs is absorbed by the cosmic gas. During this
epoch the $y$-parameter is
\begin{equation}
y_1={1 \over 4 U(z_{eq})} \int_{z_{eq}}^{z_{rec}} {dz \over aH(z)}{d\Delta U(z) \over dt}
\label{eq:phase1}
\end{equation}
where $d \Delta U(z)/dt$ is the total energy per unit comoving volume
per unit time emitted by accretion onto PBHs and $U(z)$ is the
unperturbed comoving energy density of the CMB at redshift $z$. We
will show that this contribution to the total $y$-parameter is
dominant. We integrate $d \Delta U(z)/dt$ starting at the redshift
of matter-radiation equality because before $z_{eq}$ the effective
cosmic sound speed approaches the value $c_s \sim c/\sqrt{3}$ and the
gas accretion rate onto PBHs is negligible.

\noindent
2) Between last scattering and decoupling the estimate of the
$y$-parameter is complicated by the fact that only a fraction of
the energy emitted by PBHs is absorbed by the gas and exchanged with
the CMB radiation.  As a zero-th order approximation we can use
\eq~(\ref{eq:phase1}), integrating between $z_{dec}$ and $z_{rec}$, to
estimate an upper limit for the $y$-distortion. The integration shows
that for PBH masses $M_{pbh}< 10^3$~M$_\odot$, $y_2 \ll y_1$ even when
we assume that all the energy is absorbed by the gas.  In
\S~\ref{ssec:temp_hii} we estimated that only a fraction $C \sim 1.4
\times 10^{-4}$ of the radiation emitted by PBHs is deposited into
heat. A more realistic estimate of $\Delta U$ is obtained by
multiplying our upper limit for $y_2$ by $C$. Hence, we do not need to
worry further about getting a more precise estimate of $y_2$ for PBH
of any mass as $y_2 \ll y_1$.

\noindent
3) After decoupling Compton heating/cooling becomes negligible and we
can calculate the $y$-parameter using the relationship:
\begin{equation}
y_3={k_B \over m_ec^2} \int_{z_{dec}}^{10} (T_e-T_{cmb}) d\tau_e.
\end{equation}
Let's estimate the upper limit for $y_3$. Assuming $T_e = {\rm
constant}  \gg T_{cmb}$ we have $y_3=1.726\times 10^{-6} (T_e/10^4~{\rm
K}) \Delta \tau_e$. In all the models consistent with WMAP3 data the
gas temperature is $T_e < 1000$~K and $\Delta \tau_e$ between redshift
$z \sim 10$ and $z \sim 100$ is $<0.05$. It follows that the upper
limit on $y_3$ is $8.6 \times 10^{-9}$ which is negligible when
compared to the FIRAS upper limit.  Finally, the value of the
$y$-parameter is the sum of the $y$ during each epoch. Thus, in our
case, we have $y \approx y_1$.

Let's now estimate $y_1$.  The total energy density emitted per unit
time per unit comoving volume is $\Delta U/dt=lL_{Ed}n_{pbh}$. Thus we
find
\begin{equation}
y_1={L_{Ed} \rho_{crit} \Omega_{dm} \over 4 M_{pbh}a_RT_0^4(1+z_{eq})}
f_{pbh}\int_{z_{eq}}^{z_{rec}} dz~{l(M_{pbh}, z) \over aH(z)}.
\end{equation}
Using \eq~(\ref{eq:lacc1}) for the dimensionless accretion luminosity
$l$ (see also \fig~\ref{fig:B}), we obtain the value of the
$y$-parameter as a function of $M_{pbh}$ and $f_{pbh}$. Imposing $y
\le 1.5 \times 10^{-5}$ we obtain upper limits for $f_{pbh}(M_{pbh})$
at 95\% confidence. The results are summarized in
\fig~\ref{fig:FIRAS}(left).

In summary, before the redshift of recombination gas accretion onto
PBHs with mass $<100$ M$_\odot$ is not greatly reduced by Compton
drag. Although the accretion luminosity during this epoch does not
contribute to increase $\tau_e$, the energy injection produce spectral
distortions of the CMB, increasing the value of the $y$-parameter. We
find that the existence of PBHs with masses $<100$ M$_\odot$ is best
constrained by upper limits on the $y$-parameter from FIRAS.

\section{Discussion and Summary}\label{sec:sum}

During the radiation era, mildly non-linear perturbations with $\delta
\rho/\rho \sim 0.5-1$ can collapse directly into primordial black
holes (PBHs). Such black holes may have masses ranging from the Planck
mass to a million solar masses, depending on the redshift of their
formation and the details of the formation mechanism.  The abundance
of evaporating PBHs with masses $< 10^{15}$~g is constrained by
observations to be a fraction $\beta \simlt 10^{-22}$ of the mean
energy density of the universe at the time of their formation, but the
existence of PBHs with masses larger than $10^{15}$~g is poorly
constrained. It is not ruled out that the bulk of the dark matter may
be composed of PBHs with masses in the range between $10^{15}$~g and
$10^{26}$~g or Planck mass relics with $M \sim 10^{-5}$ g.  In this
work, the last of a series of three papers, we study the effects of a
(yet undetected) population of non-evaporating PBHs on the thermal and
ionization history of the universe and their signatures on the CMB
anisotropies and spectrum.  In Paper~I we focused on studying the
formation and growth of the dark matter halo which envelopes PBHs that
do not constitute the bulk of the dark matter.  In the second paper
\citep{Ricotti:06}, we study in detail the Bondi-type accretion
solutions onto PBHs including the effects of Compton drag, Hubble
expansion and the growth of the dark matter halo. Finally, this work
focuses on modeling the accretion luminosity of PBHs including
feedback effects and observational signatures.
\begin{figure*}[t]
\epsscale{1.1}
\plottwo{\figname{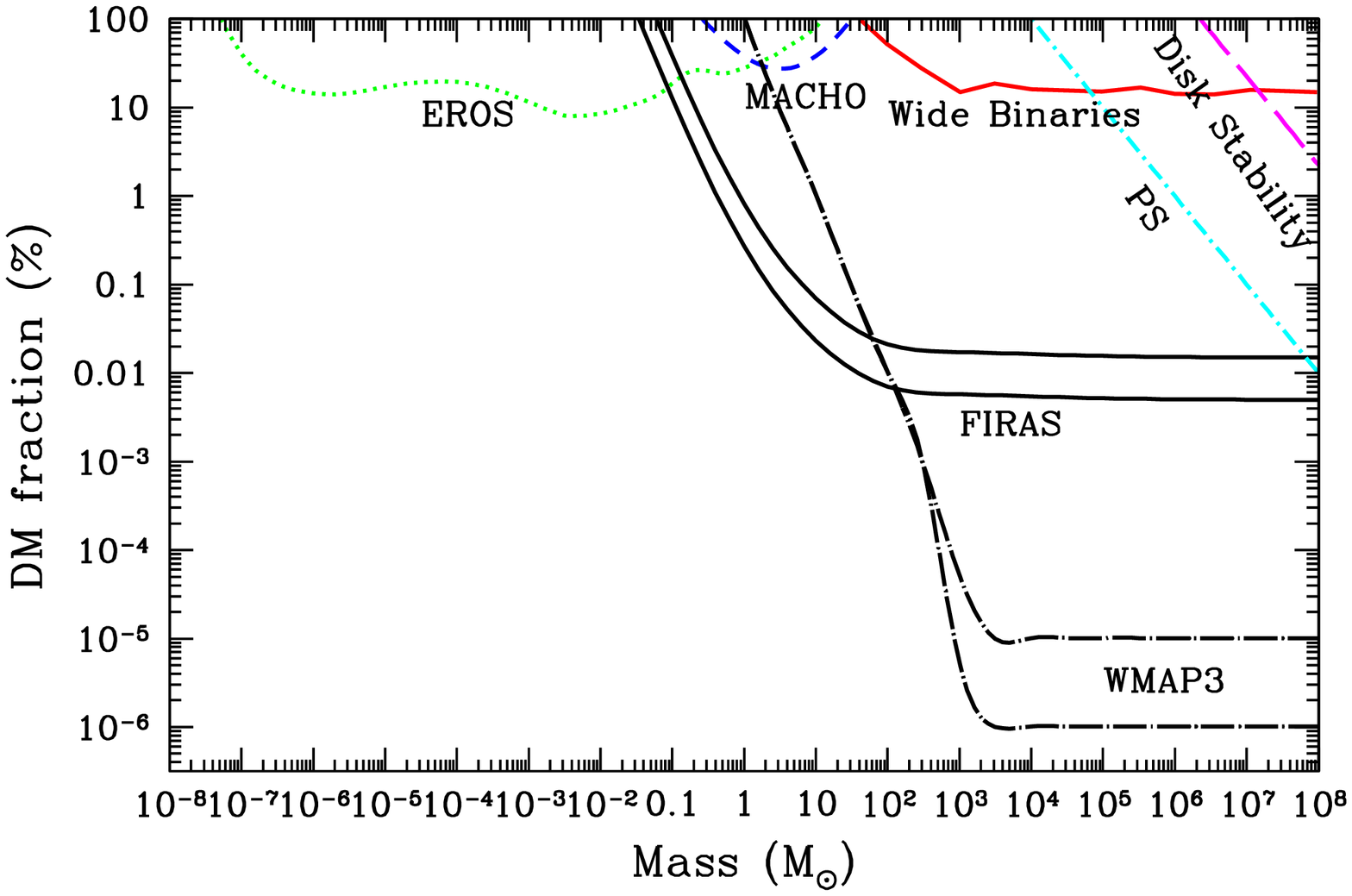}}{\figname{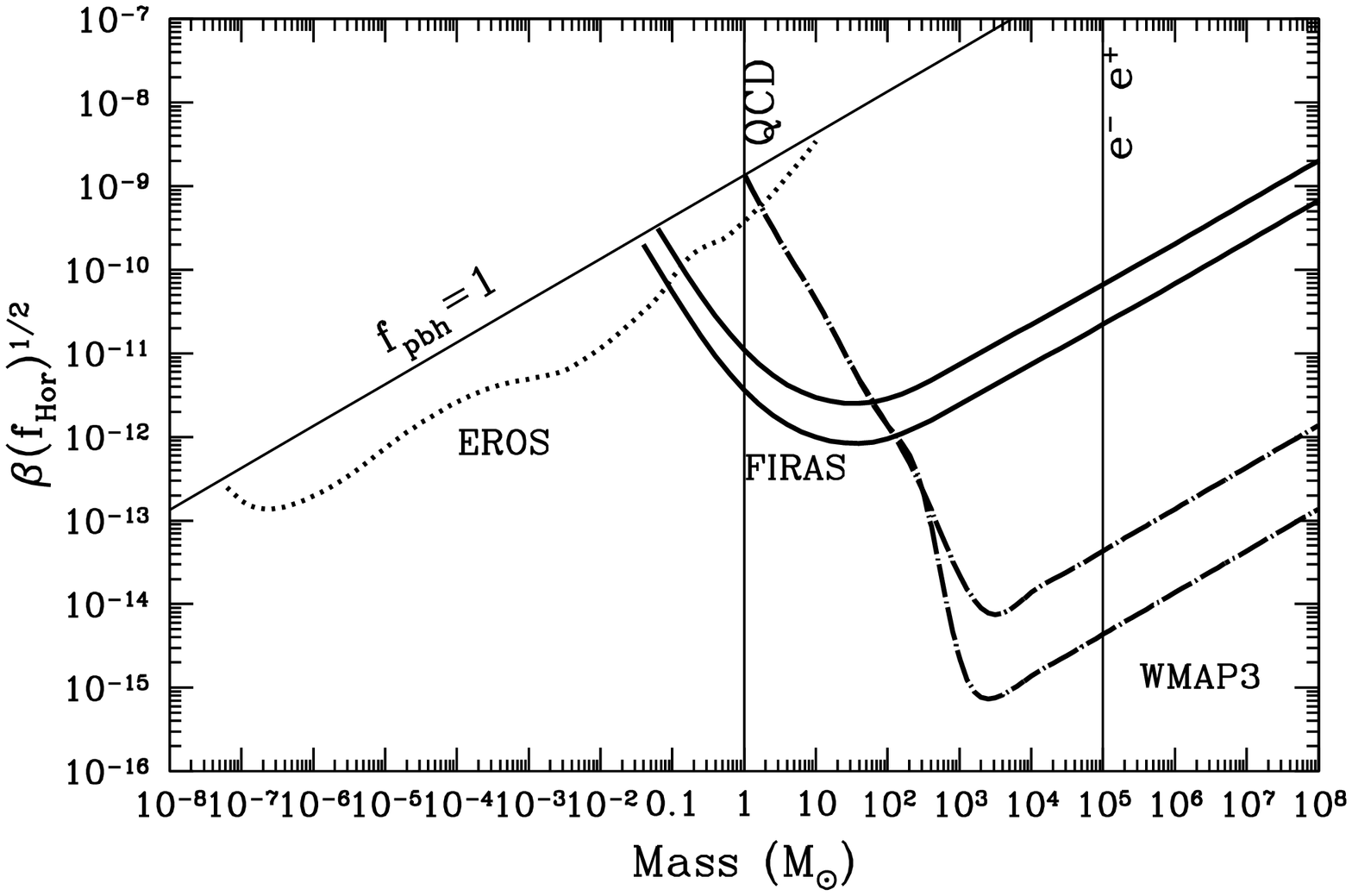}}
\caption{(left) Upper limits on the present abundance of PBHs. The
  thick lines are the results obtained in the present work. The solid
  lines show the upper limits using WMAP3 data (CMB anisotropies) for
  two values of the black hole duty cycle $f_{duty}=1$ and $0.1$. The
  dashed lines show the limits using FIRAS data (CMB spectral
  distortions) at 95\% and 68\% confidence. The other lines refer to
  previous upper limits from microlensing (EROS and MACHO experiments)
  and dynamical constraints (see introduction). (Right) Upper limits on
  the abundance of PBHs at the epoch of their formation $\beta$ as a
  function of their mass. We assume that the mass of PBHs is a
  fraction $f_{Hor}$ of the mass of the horizon at the epoch of their
  formation. The thick curves show the upper limits obtained in the
  present work and the thin dotted curve are limits from the EROS
  collaboration (microlensing experiment).}
\label{fig:FIRAS}
\end{figure*}

We find that if a fraction $f_{pbh}$ of the dark matter is in PBHs
with mass $> 0.1$~M$_\odot$, the energy released due to gas accretion
may produce spectral distortions of the CMB radiation and keep the
universe partially ionized after recombination.  The limits on the
mass and abundances of PBHs set from observations of the X-ray
background are much less restrictive than those from the CMB.  The
modified recombination history produces observable signatures on the
spectrum of polarization anisotropies of the CMB at angular scales $l
\simgt 10$. Hence, the effect of PBHs cannot be confused with the
effect of ionization by high redshift galaxies which affect
polarization anisotropies on larger angular scales.

We are able to improve the constraints on $f_{pbh}$ for PBHs with
masses $> 0.1$~M$_\odot$ by several orders of magnitude using WMAP3
and FIRAS data. The results are summarized in \fig~\ref{fig:FIRAS}
(left).  The upper limits on the abundance of PBHs with masses $0.1
M_\odot < M_{pbh}<10^8$ M$_\odot$ at the epoch of their formation,
$\beta$, are shown in \fig~\ref{fig:FIRAS} (right). We use
\eq~(\ref{eq:relic}) to derive $\beta$ as a function of $M_{pbh}$,
$f_{pbh}$ and the ratio $f_{Hor}=M_{pbh}/M_H$ between the PBH mass and
the mass $M_H$ of the horizon at the epoch of PBH formation.

Fitting WMAP3 data with cosmological models that do not allow for
non-standard recombination histories as produced by PBHs or other
early energy sources may lead to an underestimate of the best-fit values of
the amplitude of linear density fluctuations, $\sigma_8$, and the
scalar spectral index, $n_s$.  This happens because the contribution of
PBHs to the optical depth to Thompson scattering, which is uncorrelated
with the contribution from galactic ionization sources, can be $\Delta
\tau_e \simlt 0.05$ ($\Delta \tau_e \simlt 0.1$ at 95\% CF). Since
$n_s$ and $\sigma_8$ are correlated with $\tau_e$, their best fit values
may increase to $n_s \sim 1$ and $\sigma_8 \sim 0.9$ (at 95\%
CF). This is a general result that may reduce recent tensions between
WMAP3 data and clusters data on the value of $\sigma_8$
\citep{Evrard:07}.

A population of intermediate mass black holes (IMBHs) with masses of
about $100-1000$ M$_\odot$ is still allowed and may be widespread if a
fraction of the ultraluminous X-ray sources (ULX) observed in nearby
galaxies host IMBHs \citep{Miller:03, Dewangan:06}. The origin of
IMBHs is unknown, but if they are produced by Pop~III stars their
number may fall short in explaining the observed ULXs population
\citep[\eg,][]{Kuranov:07, Pelupessy:07}.
We find that, if all or a fraction of observed ULXs are PBHs with
masses $M_{pbh}\sim 100-1000$~M$_\odot$ with $f_{pbh}\sim 10^{-5}$,
they would increase the best fit value of the optical depth to
Thompson scattering to $\tau_e \sim 0.2$. Since the scalar spectral
index $n_s$ and the amplitude of density fluctuations $A_s$ and
$\sigma_8$ are correlated to $\tau_e$, their best fits also increase to
$n_s \sim 1$ and $\sigma_8 \sim 0.9$. PBHs in this mass range may be
produced in two-stage inflationary models designed to fit the low WMAP
quadrupole \citep{Kawasaki:06}. We emphasize again that this effect is
more general than the specific case of PBHs discussed in this
paper. Any mechanism or energy source that modifies the standard
recombination history may affect the estimate of cosmological
parameters in a way similar to that discussed here.

Our results are in contradiction with the suggestion that MACHOs are
PBHs with mass $\sim 0.1-1$~M$_\odot$ and $f_{pbh}\sim 0.2$
\citep{Alcock:00}. Such a PBH population would produce spectral
distortions incompatible with FIRAS data.

The luminous QSOs found by SLOAN at $z \sim 6$ are thought to be
powered by $10^8-10^9$ M$_\odot$ SMBHs. It is difficult to produce
such massive black holes starting from small seeds by gas accretion
because the age of the universe at $z=6$ is a few tens the Salpeter
accretion timescale.  A few massive PBHs or numerous less massive PBHs
may help explain the origin of SMBHs at high redshift and in present
day galaxies by producing relatively massive ``seeds''. Are the upper
limits on the number of PBHs derived in this work compatible with this
scenario? The fraction of mass in SMBHs today is approximately
$\Omega_{smbh}/\Omega_{dm} \sim 2.13 \times 10^{-5}$
\citep{Gebhardt:00, RicottiO:03}. For PBHs with mass $>1000$ M$_\odot$
we found $f_{pbh}=\Omega_{pbh}/\Omega_{dm} \simlt
10^{-6}/f_{duty}$. Hence, assuming that only a fraction $F_{agn}\le 1$
of PBHs is incorporated into SMBHs and grows by gas accretion by a
factor $X_{acc} \ge 1$ we have: $f_{pbh}X_{acc}F_{agn} \sim 2 \times
10^{-5}$ or $X_{acc}F_{agn} \simgt 20f_{duty}$. The most massive PBHs
have $F_{agn} \rightarrow 1$ because they spiral in to the centers of
galaxies by dynamical friction on a shorter timescale
($t_{fric}/t_H(z) \sim 0.02 M_{halo}(z)/M_{pbh}$, where $t_H$ is the
Hubble time) and because they may accrete gas more efficiently.
Hence, for $f_{duty}\sim 3\%$ and $F_{agn}=1$ we find $X_{acc} \simgt
1$ indicating that even scenarios with negligible mass accretion onto
PBHs (\ie, only growth through mergers) are consistent with the
observed mass in SMBHs today.

Less massive PBHs have lower probability for growing to masses typical
of SMBHs because the Bondi accretion rate is $\propto M^2$. However,
the upper limit on the abundance of PBHs increases steeply with
decreasing mass for $M_{pbh} <1000$ M$_\odot$. Thus, although a
smaller fraction of the seed PBHs can grow substantially, the number
of seeds available can be much larger. PBHs with masses smaller than
$100$ M$_\odot$, assuming Bondi type accretion from the ISM of a
typical high-z galaxy, are unlikely to accrete rapidly enough to grow
to SMBH masses in less than 1 Gyr, even if they constitute a few per
cent of the dark matter \citep{Kuranov:07, Pelupessy:07, RicottiK:07}.

The increased fractional ionization of the cosmic gas produced by
non-standard recombination also increases the primordial molecular
hydrogen abundance to $x_{H_2} \sim 10^{-4}-10^{-5}$ after redshift $z
\sim 100$. This value is between ten and one hundred times larger than
the standard value, $x_{H_2} \sim 10^{-6}$, obtained neglecting
PBHs. The increase of the cosmic Jeans mass due to X-ray heating is
negligible for models consistent with the CMB data. Therefore, the
formation rate of the first galaxies and stars may be enhanced if a
population of PBHs exists.  Several aspects of first-star and galaxy-
formation physics would be affected by the enhanced molecular
fraction: (i) the mass of the first stars may be reduced due to
formation of HD molecules \citep{Nagakura:05}; (ii) the intergalactic
medium would be optically thick to H$_2$ photo-dissociating radiation
in the Lyman-Werner bands, allowing molecular hydrogen to survive in the
low density IGM even at relatively low redshifts $z \sim 10-15$; (iii)
the epoch of domination of the first stars and galaxies would probably
start earlier and perhaps last longer. The number of first galaxies
that remain completely dark would be reduced. It is not obvious that
the star formation efficiency and other internal properties of the
first galaxies would be affected because feedback effects such as
photo-evaporation from internal sources and SN explosions are probably
dominant \citep{RicottiGSa:02, RicottiGSb:02}.  We leave quantitative
calculations on the impact of PBHs on the formation of the first
galaxies to a future work.

\subsection*{ACKNOWLEDGMENTS}

MR acknowledge stimulating discussions with Niayesh Afshordi,
Carlo Burigana, Andrea Ferrara, Avi Loeb, Cole Miller, Eve Ostriker,
Ruben Salvaterra, David Spergel and Mathias Zaldariaga. This work was
supported in part by NASA grant NNX07AH10G (MR) and a NSF Graduate
Research Fellowship (KJM).


\bibliographystyle{/home/ricotti/Latex/TeX/apj}
\bibliography{/home/ricotti/Latex/TeX/archive}

\label{lastpage}
\end{document}